\newcommand{\beq}{\begin{equation}}
\newcommand{\eeq}{\end{equation}}

\newcommand{\bear}{\begin{eqnarray}}
\newcommand{\eear}{\end{eqnarray}}

\documentclass[twoside,twocolumn]{article}

\usepackage{amsmath}
\usepackage{amssymb}
\usepackage[sc]{mathpazo} 
\usepackage[T1]{fontenc} 
\usepackage{microtype}
\usepackage[english]{babel}
\usepackage{graphicx}
\usepackage[titletoc,title]{appendix}

\usepackage[left=48pt,right=42pt,top=46pt,bottom=60pt,headheight=15pt,headsep=10pt,letterpaper,twoside]{geometry}%
\usepackage[labelfont={bf,sf},labelsep=period,figurename=Fig.,singlelinecheck=off,          justification=RaggedRight]{caption}

\usepackage{booktabs} 

\usepackage{enumitem} 
\setlist[itemize]{noitemsep} 

\usepackage{abstract} 

\usepackage{titlesec} 
\renewcommand\thesection{\Roman{section}} 
\renewcommand\thesubsection{\roman{subsection}} 
\titleformat{\section}[block]{\large\scshape\centering}{\thesection.}{1em}{} 
\titleformat{\subsection}[block]{\large}{\thesubsection.}{1em}{} 

\usepackage{fancyhdr} 
\usepackage{titling} 

\usepackage{hyperref} 

\title{A new imaging technology based on Compton X-ray scattering} 
\author{%
\'Angela Sa\'a Hern\'andez\thanks{Corresponding author:  angela.saa.hernandez@usc.es},  Diego Gonz\'alez-D\'iaz, Marcos Seoane \\[1ex]
\normalsize Instituto Galego de F\'isica de Altas Enerx\'ias (IGFAE)\\
\normalsize R\'ua de Xoaqu\'in D\'iaz de R\'abago, s/n, Campus Vida, 15782 Santiago de Compostela, Spain \\
\\
Carlos Azevedo \\[1ex] 
\normalsize I3N, Physics Department, University of Aveiro\\
\normalsize Campus Universit\'ario de Santiago, 3810-193 Aveiro, Portugal \\
\\
Pablo Villanueva \\[1ex] 
\normalsize Department of Physics, Lund University\\
\normalsize P.O. Box 118, SE-22100 Lund, Sweden \\
}
\date{} 
\renewcommand{\maketitlehookd}{%
\begin{abstract}
We describe a feasible implementation of a novel X-ray detector for highly energetic x-ray photons with a large solid angle coverage, optimal for the detection of Compton x-ray scattered photons. The device consists of a 20~cm-thick sensitive volume filled with xenon at atmospheric pressure. When the Compton-scattered photons interact with the xenon, the released photoelectrons create clouds of secondary ionization, which are imaged using the electroluminescence produced in a custom-made multi-hole acrylic structure. Photon-by-photon counting can be achieved by processing the resulting image, taken in a continuous readout mode. Based on Geant4 simulations, by considering a realistic detector design and response, we show that photon rates up to at least $10^{11}$ ph/s on-sample ($5~{\mu}$m water-equivalent cell) can be processed, limited by the spatial diffusion of the photoelectrons in the gas. Illustratively, if making use of the Rose criterion and assuming the dose partitioning theorem, we show how such a detector would allow obtaining 3d images of $5~{\mu}$m-size unstained cells in their native environment in about 24~h, with a resolution of 36~nm.
\end{abstract}
}

\begin{document}

\maketitle

\section{Introduction}

\begin{figure}[h!!!]
    \centering
    \includegraphics[width=82mm]{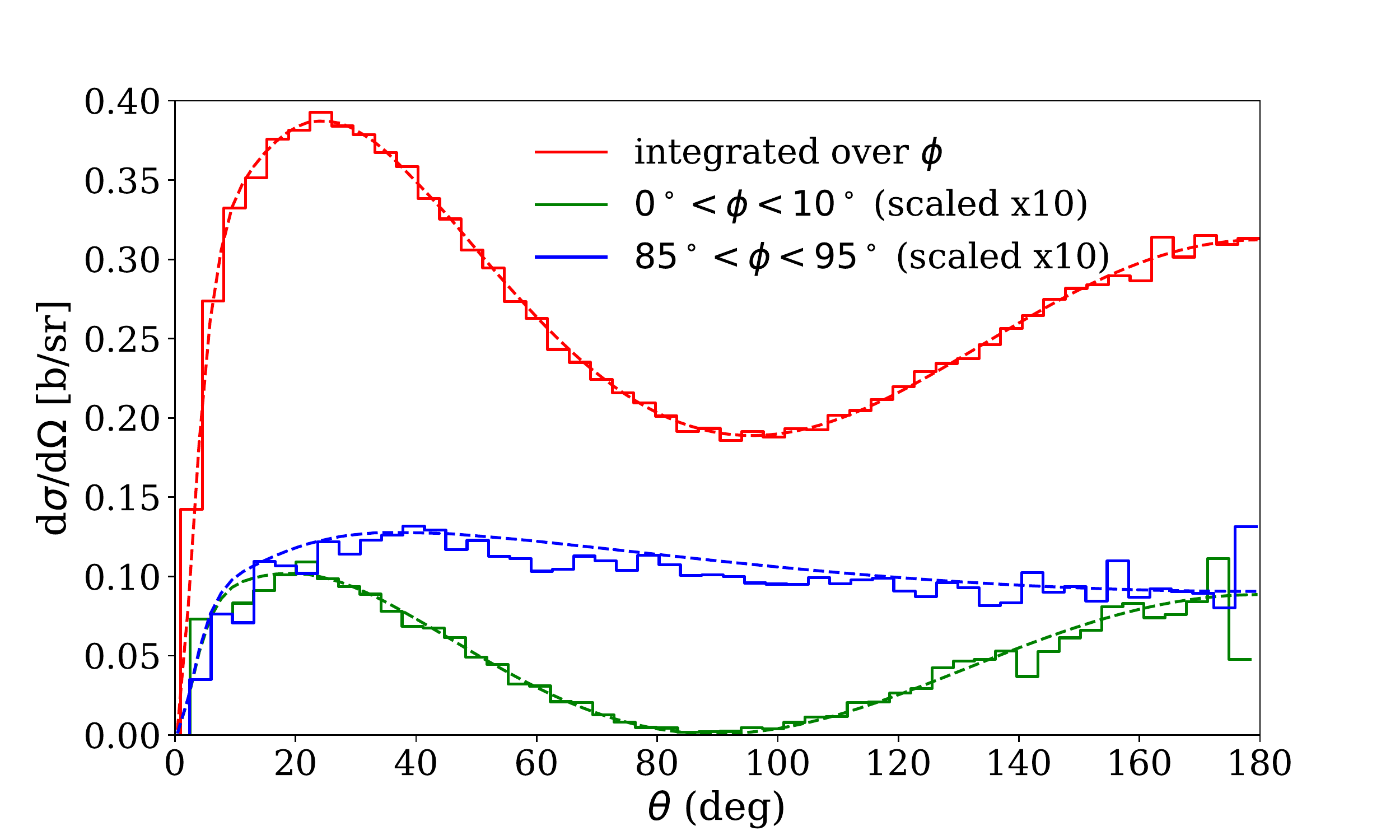}
    \caption{Differential cross section for Compton-scattered photons on DNA (in barn per stereoradian), for a linearly polarized x-ray beam of 64~keV as obtained with Monte Carlo simulations (using Geant4~\cite{Geant4}) and tabulated values~\cite{Hub75} (dashed lines), for different azimuthal regions: $\phi=[0-10]^\circ$(green), $\phi=[85-95]^\circ$(blue) and integrated over $\phi$ (red). $\phi$ indicates the angle relative to the direction of the polarization vector.}
    \label{fig:Angular_distribution}
\end{figure}

Despite some x-ray facilities and experiments make use of Compton scattering to probe for instance the electronic and magnetic structure of materials~\cite{Sak98, Tsc98}, the limited flux and brilliance (brightness) that is currently available at the required high energies ($\gtrsim$20~keV), seem to have precluded the popularization of these techniques.
With the advent of the 4$^{\scriptsize\textnormal{th}}$ generation of synchrotron light sources, such as ESRF-EBS~\cite{ESRF}, the projected APS-U~\cite{APS}, Petra~IV~\cite{PETRA}, and SPring-8-II~\cite{SPring8}, as well as the proposal of novel facilities based on x-ray free-electron lasers~\cite{Hua13}, which increase the brightness and coherent flux for hard x-rays at least two orders of magnitude beyond today’s capability, a unique opportunity arises to use Compton scattering in ways that were not conceived before.
An example of these new possibilities is scanning Compton x-ray microscopy (SCXM)~\cite{Vil18}. This technique has the potential of obtaining 10's of nanometer resolution images of biological or radiosensitive samples without sectioning or labelling. Thus, it bridges the capabilities of optical and electron microscopes.
Exploiting Compton interactions for biological imaging is possible because, in spite of its inelastic nature, the SCXM technique makes an optimal use of the number of scattered photons per unit dose, i.e., the deposited energy per unit of mass.
Generally speaking, an efficient use of Compton scattering implies, first and foremost, that a nearly 4$\pi$-coverage is required (Fig.~\ref{fig:Angular_distribution}), at an optimal energy around 64~keV if aiming for instance at resolving DNA structures \cite{Vil18}. This poses a formidable challenge for current detection technologies, which are costly and have detection areas much below the required size. Conversely, at lower x-ray energies ($\lesssim$10~keV), imaging based on coherent scattering has benefited from the development of ultra-fast pixelated silicon detectors, capable of performing photon-counting up to $10^7$ counts/s/pixel. A nowadays typical detection area is $40\times40$ cm$^2$, sufficient for covering the coherent forward cone at a distance of about 1~m, at near 100\% quantum efficiency \cite{HPC_review}. At higher energies, silicon must be replaced by a semi-conductor with a higher stopping power to x-rays, e.g., CdTe. However, targeting a geometrical acceptance around $70\%$ at 64~keV, while providing enough space to incorporate a compact setup (namely the sample holder, step motor, pipes, shielding and associated mechanics), would imply an imposing active area for these type of detectors, well above 1000~cm$^2$. For comparison, PILATUS3 X CdTe 2M, one of the latest high-energy x-ray detectors used at synchrotron sources, has an active area of 25$\times$28~cm$^2$ \cite{Pilatus}.
Clearly, the availability of a 4$\pi$/high energy x-ray detector would soon become an important asset at any next generation facility, if it can be implemented in a practical way.

In this work we have implemented a novel approach for the detection of 4$\pi$ Compton-scattered photons based on a technology borrowed from particle physics:
the electroluminescent Time Projection Chamber (EL-TPC), discussing its performance as an SCX-microscope. TPCs, introduced by D. Nygren in 1974 \cite{Nyg74, Nyg18} are nowadays ubiquitous in particle and nuclear physics, chiefly used for reconstructing particle interactions at high track multiplicities \cite{ALICE}, and/or when very accurate event reconstruction is needed \cite{DDM-Lomba, DUNE, Gon18}. The main characteristics of the particular TPC-flavour proposed here can be summarized as: i) efficient to high energy x-rays thanks to the use of xenon as the active medium, ii) continuous readout mode with a time sampling around ${\Delta}T_s = 0.5~{\mu}$s, iii) typical temporal extent of an x-ray signal (at mid-chamber): ${\Delta}T_{x-ray} = 1.35~{\mu}$s, iv) about 2000 readout pixels/pads, v) single-photon counting capability, and vi) an energy resolution potentially down to 2\% FWHM for 60~keV x-rays, thanks to the electroluminescence mode~\cite{Fano}, only limited by the Fano factor $F$.\footnote{A non-zero value of $F$ stems from the the intrinsic spread of primary ionization, as the partition of energy between excitations and ionizations changes event by event.} Importantly, the distinct advantage of using electroluminescence instead of conventional avalanche multiplication is the suppression of ion space charge, traditionally a shortcoming of TPCs operated under high rates.

Our design is inspired by the proposal in \cite{Dave_prop}, that has been successfully adopted by the NEXT collaboration in order to measure neutrino-less double-beta decay \cite{Francesc}, but we include three main simplifications: i) operation at atmospheric pressure, to facilitate the integration and operation at present x-ray sources, ii) removal of the photomultiplier-based energy-plane, and iii) introduction of a compact all-in-one electroluminescence structure, purposely designed for photon-counting experiments.

In this paper we discuss, starting from section~\ref{sectDesign}, the main concepts and working principles leading to our conceptual detector design. Next, in section \ref{TPCresponse}, we study the photon counting capabilities of a realistic detector implementation. We present the expected performance when applied to the SCXM technique in section~\ref{results}. Finally, we assess the limits and scope of the proposed technology in section~\ref{discussion}.

\section{TPC design} \label{sectDesign}
\subsection{Dose and intrinsic resolving power}

\begin{figure}[t]
    \centering
    \includegraphics[width=82mm]{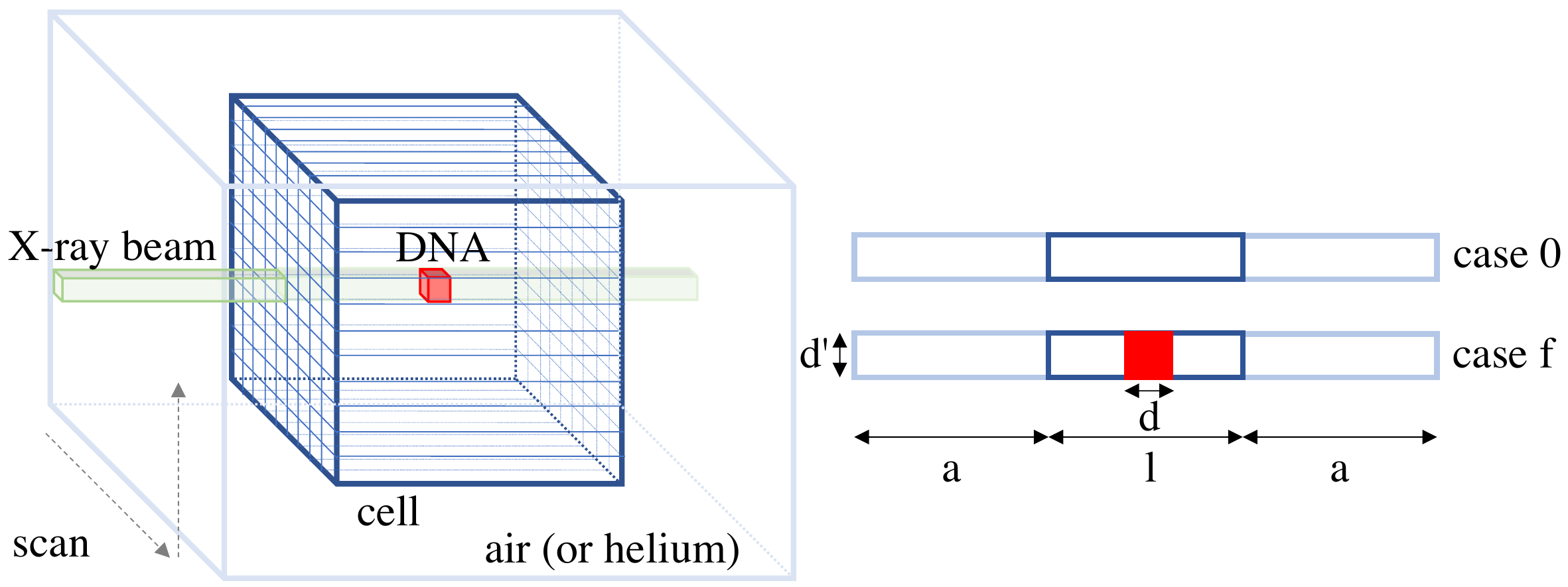}
    \includegraphics[width=82mm]{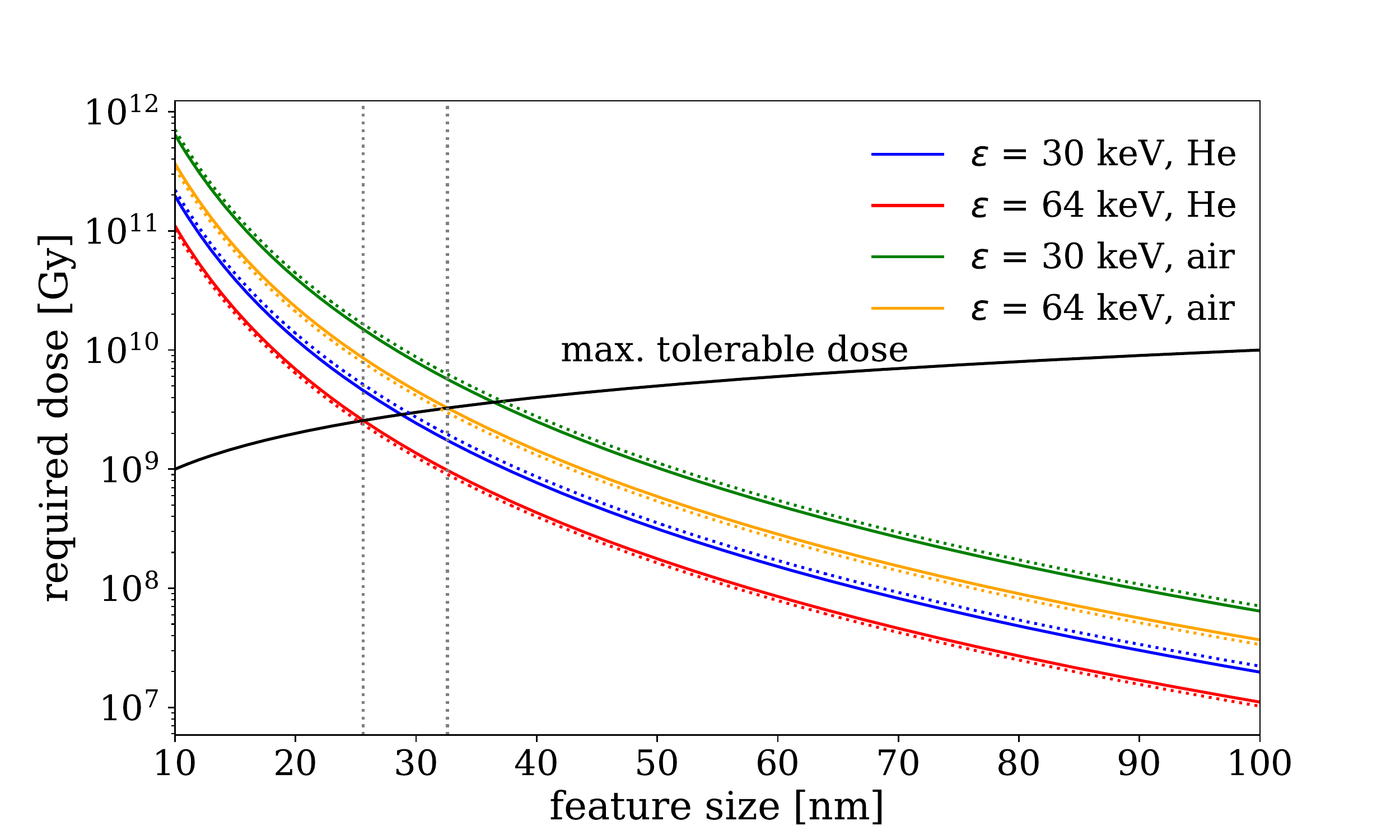}
    \caption{Top: study case. A cubic DNA feature (size $d$) is embedded in a cubic water cell ($l= 5~{\mu}$m), surrounded by air/helium ($a= 5$ mm). The photon beam scans regions containing only water (case 0), or water and DNA (case f). These two cases are used to evaluate the resolving power of SCXM at a given dose.
    Bottom: dose needed to resolve a DNA feature as a function of its size assuming 100\% detection efficiency, for x-ray energies of 30~keV and 64~keV, obtained respectively with Geant4~\cite{Geant4} (solid lines) and using NIST values \cite{NIST} (dotted line), and the formulas in text. The black line represents the maximum tolerable dose estimated from coherent scattering experiments ~\cite{ChapmanDose}.}
    \label{fig:Dose}
\end{figure}

In a scanning, dark-field, configuration, the ability to resolve a feature of a given size embedded in a medium can be studied through the schematic representation shown in Fig. \ref{fig:Dose}-top, that corresponds to an arbitrary step within a 2d-scan, in a similar manner as presented in~\cite{Vil18}.

Three main assumptions lead to this simplified picture: i) the dose fractionation theorem \cite{DoseFrac}, based on which one can expect 3d reconstruction capabilities at the same resolution (and for the same dose) than in a single 2d-scan, ii) the ability to obtain a focal spot, $d'$, down to a size comparable to (or below) that of the feature to be resolved, $d$, and iii) a depth of focus exceeding the dimensions of the sample under study, $l$.
We adopt the situation in Fig. \ref{fig:Dose}-top as our benchmark case, and we use the Rose criterion \cite{Rose} as the condition needed to discern case $f$ (feature embedded within the scanned volume) from case $0$ (no feature), that reads in the Poisson limit as:
\begin{equation}
\frac{|N_{f} - N_{0}|}{\sqrt{\sigma_{N_{f}}^2 + \sigma_{N_{0}}^2}} =\frac{|N_{f} - N_{0}|}{\sqrt{N_{f} + N_{0}}} \geq 5   \label{Rose}
\end{equation}
with $N$ being the number of scattered photons. Substitution of physical variables in eq. \ref{Rose} leads directly to a required fluence of:
\begin{equation}
\phi \geq \phi_{min} = 25\frac{(2l-d)\!\cdot\!\lambda^{-1}_w+d\!\cdot\!\lambda^{-1}_f+4\!\cdot\!a\!\cdot\!\lambda^{-1}_a}{d'^2\!\cdot\!d^2\!\cdot\!(\lambda^{-1}_f-\lambda^{-1}_w)^2} \label{PhiMin}
\end{equation}
and we will assume $d'\simeq d$. Here $\lambda_w$, $\lambda_f$, $\lambda_a$ are the Compton-scattering mean free paths of x-rays in water, DNA, and air (or helium), respectively (table~\ref{tab:material_parameters}), and dimensions are defined in Fig. \ref{fig:Dose}-top. Finally, we evaluate the dose that will be imparted at the feature in these conditions as:
\begin{equation}
\mathcal{D} \! = \! \phi_{min}\!\cdot\!\varepsilon\!\cdot\!\frac{N_A}{M_f}\!\!\cdot\!\!\!\left[\sigma_{ph}\! +\!\! \int\!\!\frac{d\sigma_{_C}}{d\Omega}\!\!\cdot\!\!(1\!-\!\frac{1}{1\!\!+\!\frac{\varepsilon}{m_ec^2}(1\!-\!\cos\theta)}\!) d\Omega \right] \label{surDos}
\end{equation}
where $\sigma_{ph}$ is the photoelectric cross section and $d\sigma_{_C}/d\Omega$ is the differential cross section for Compton scattering, both evaluated at the feature. $M_f$ is the feature molar mass, $N_A$ the Avogadro number, $\varepsilon$ the photon energy and $\theta$ its scattering angle. The dose inherits the approximate $l/d^4$ behaviour displayed in equation (\ref{PhiMin}).

\begin{table}[h]
    \caption{Mean free path for different materials at the studied energies 30 and 64~keV, according to NIST.}
    \begin{tabular}{p{0.16\textwidth}p{0.08\textwidth}p{0.08\textwidth}p{0.08\textwidth}}
    \hline
    Mean free path & 30 keV & 64 keV & Material \\
    \hline
    $\lambda_w$~[cm] & 5.47    & 5.69    & water\\
    $\lambda_f$~[cm] & 3.48    & 3.54    & DNA \\
    $\lambda_a$~[cm] & 4950.49 & 4945.60 & air \\
    \hline
    \end{tabular}
    \label{tab:material_parameters}
\end{table}

Working with eq. \ref{surDos} is convenient because it has been used earlier, in the context of coherent scattering, as a metric for assessing the maximum radiation prior to inducing structural damage~\cite{ChapmanDose}. By resorting to that estimate (black line in Fig. \ref{fig:Dose}-bottom), the doses required for resolving a feature of a given size can be put into perspective. These doses, obtained using Geant4 for a DNA feature embedded in a $5~{\mu}$m water-equivalent cell, are shown as continuous lines. Results resorting to NIST values~\cite{NIST} and Hubbell parameterization for $d\sigma_{_C}/d\Omega$ ~\cite{Hub75} are displayed as dashed lines, highlighting the mutual consistency in this simplified case. Clearly, SCXM can potentially resolve 33~nm-size DNA features inside $5~{\mu}$m cells, and down to 26~nm if a stable He atmosphere around the target can be provided.

Using eq. \ref{surDos} as a valid metric for inter-comparison between SCXM and coherent scattering is at the moment an open question and will require experimental verification. In particular, the formula assumes implicitly that the energy is released locally. However, a 10~keV photoelectron has a range of up to $2~{\mu}$m in water, while a 64~keV one can reach $50~{\mu}$m. An approximate argument can be sketched based on the fact that the average energy of a Compton electron for 64~keV x-rays (in the range 0-14~keV) is similar to that of a 10~keV photo-electron stemming from 10~keV x-rays, a typical case in coherent diffraction imaging (CDI). Given that at 64~keV most (around 70\%) of the energy is released in Compton scatters, the situation in terms of locality will largely resemble that of coherent scattering. Hence, compared to CDI, only about 30\% of the energy will be carried away from the interaction region by the energetic 64~keV photoelectrons. On the other hand, at 30~keV (the other energy considered in this study) the photoelectric effect contributes to 90\% of the dose, so one can expect a higher dose tolerance for SCXM than the one estimated here.

Naturally, the shielding pipes, the structural materials of the detector, the detector efficiency, the instrumental effects during the reconstruction, and the accuracy of the counting algorithms can limit the achievable resolution, resulting in dose values larger than the ones in Fig. \ref{fig:Dose}. These effects are discussed in the next sections.

\begin{figure}[h!!]
    \centering
    \includegraphics[width=70mm]{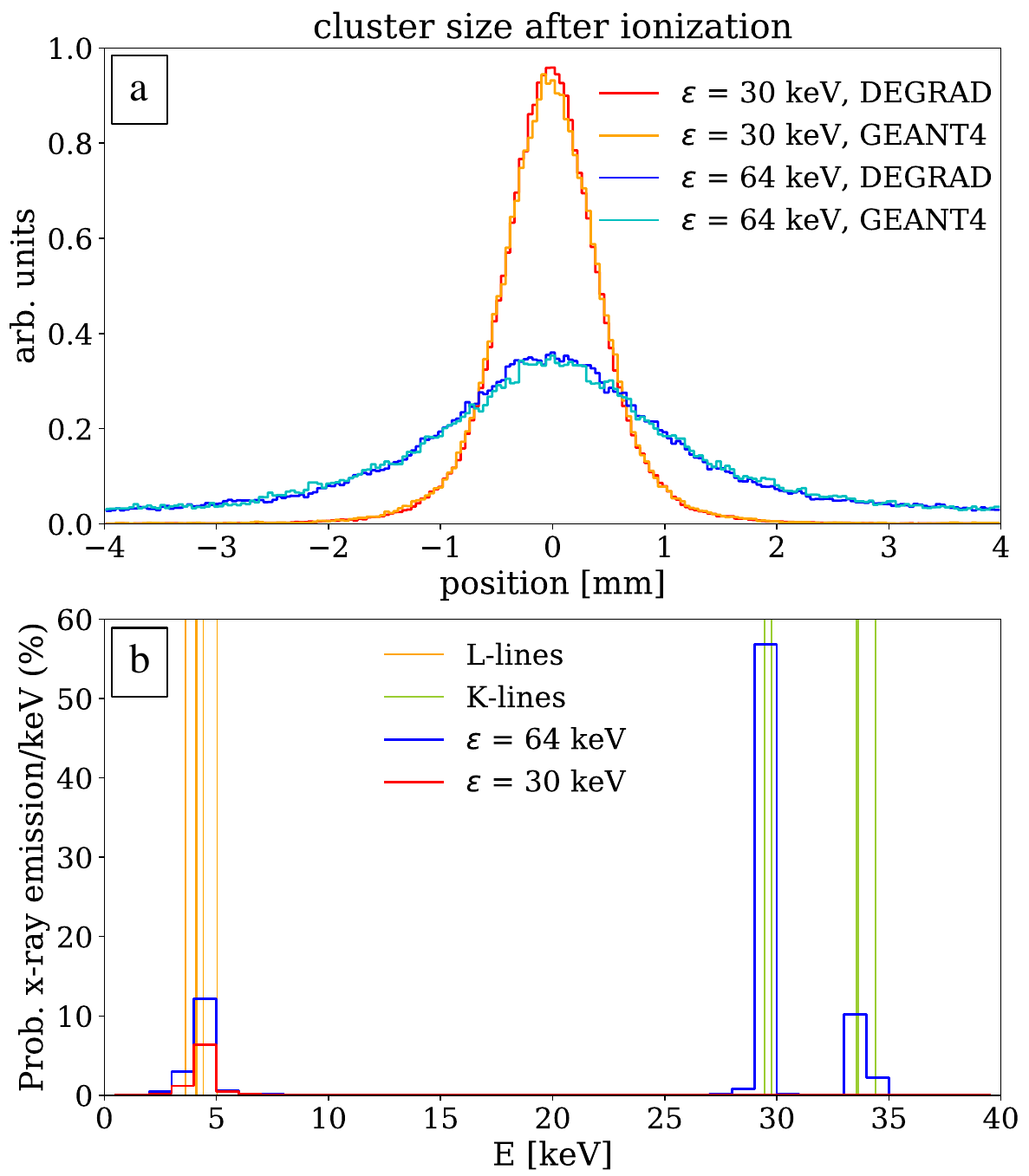}
    \includegraphics[width=71mm]{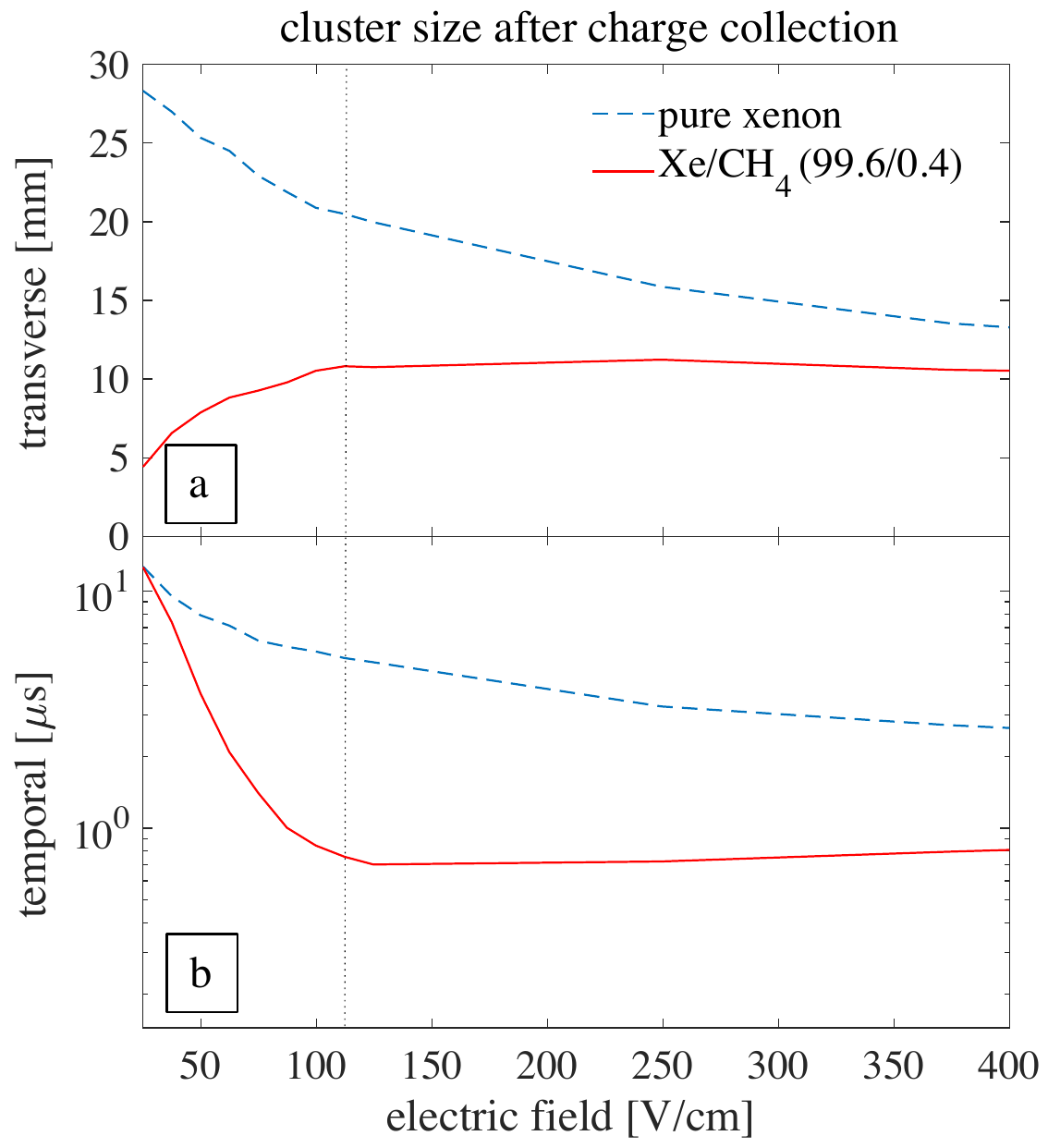}
    \caption{Top(a): ionization distributions in xenon gas, stemming from x-rays interacting in an infinite volume. They are obtained after aligning each x-ray ionization cloud by its barycenter, and projecting it over an arbitrary axis. Calculations from Geant4 are compared with the microscopic code DEGRAD developed by S. Biagi \cite{Bia13}. Top(b): probability of characteristic x-ray emission in xenon for an incident photon energy of 30~keV (red) and 64~keV (blue), in Geant4. The K-shell (green) and L-shell (orange) lines, as tabulated in~\cite{Booklet}, are shown for comparison. Bottom(a): transverse size of a point-like ionization cluster after drifting along 50~cm, obtained from Magboltz. Bottom(b): longitudinal size of a point-like ionization cluster (in time units), in the same conditions. Results for pure xenon and a fast `counting' mixture based on Xe/CH$_4$ are shown for comparison.}
    \label{fig:ClusterSize}
\end{figure}

\subsection{Technical description of the TPC working principle}

When x-rays of energies of the order of 10's of keV interact in xenon gas at atmospheric pressure, the released photoelectron creates a cloud of secondary ionization (containing thousands of electrons) with a typical ($1\sigma$) size of 0.25-1~mm (Fig. \ref{fig:ClusterSize}-top). If the x-ray energy is above that of the xenon K-shell, characteristic emission around 30-34~keV will ensue, in about 70\% of the cases. At these energies, x-ray interactions in xenon take place primarily through photoelectric effect, with just a small ($\lesssim 1\%$) probability of Compton scattering. 

\begin{figure}[h!]
    \centering
    \includegraphics[width=70mm]{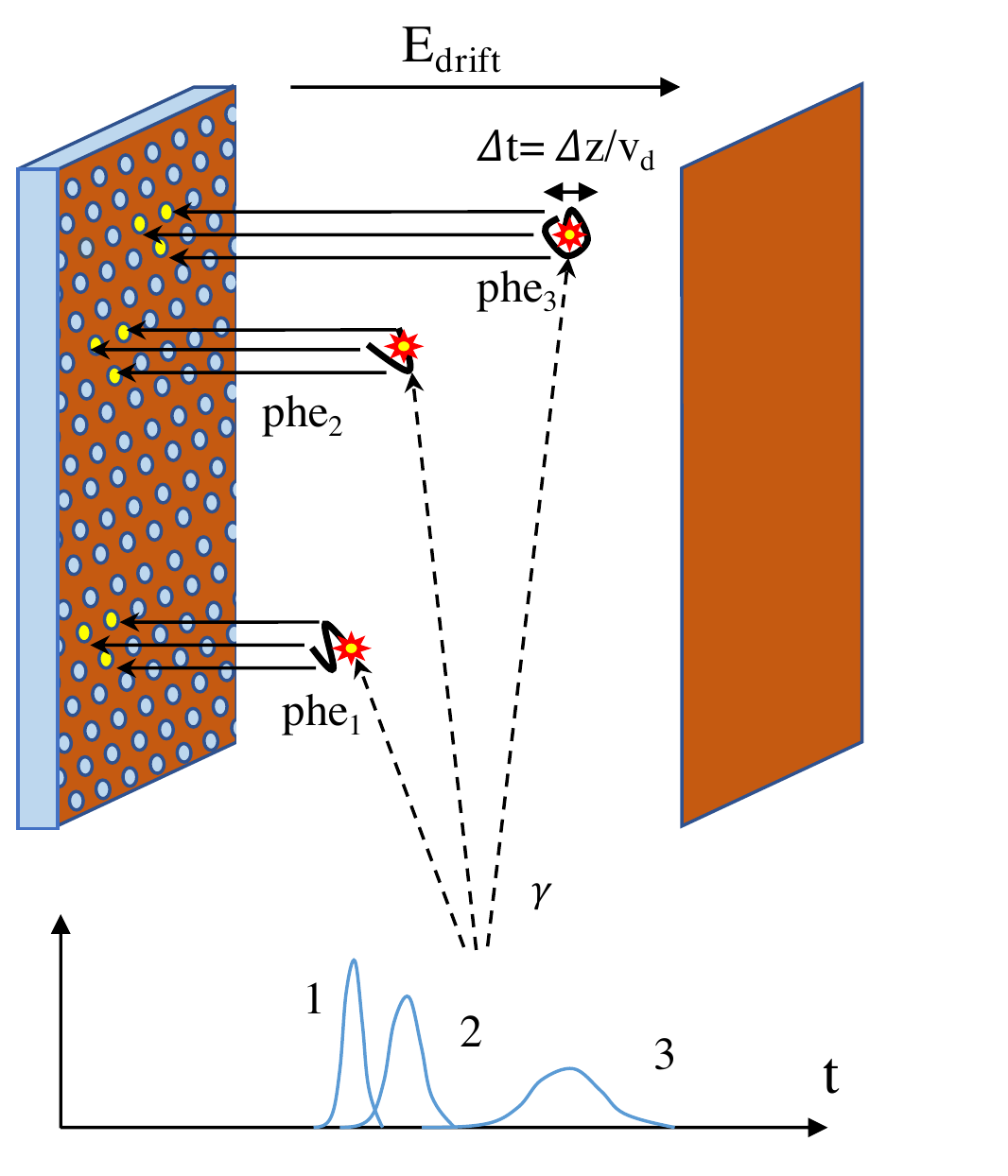}
    \includegraphics[width=80mm]{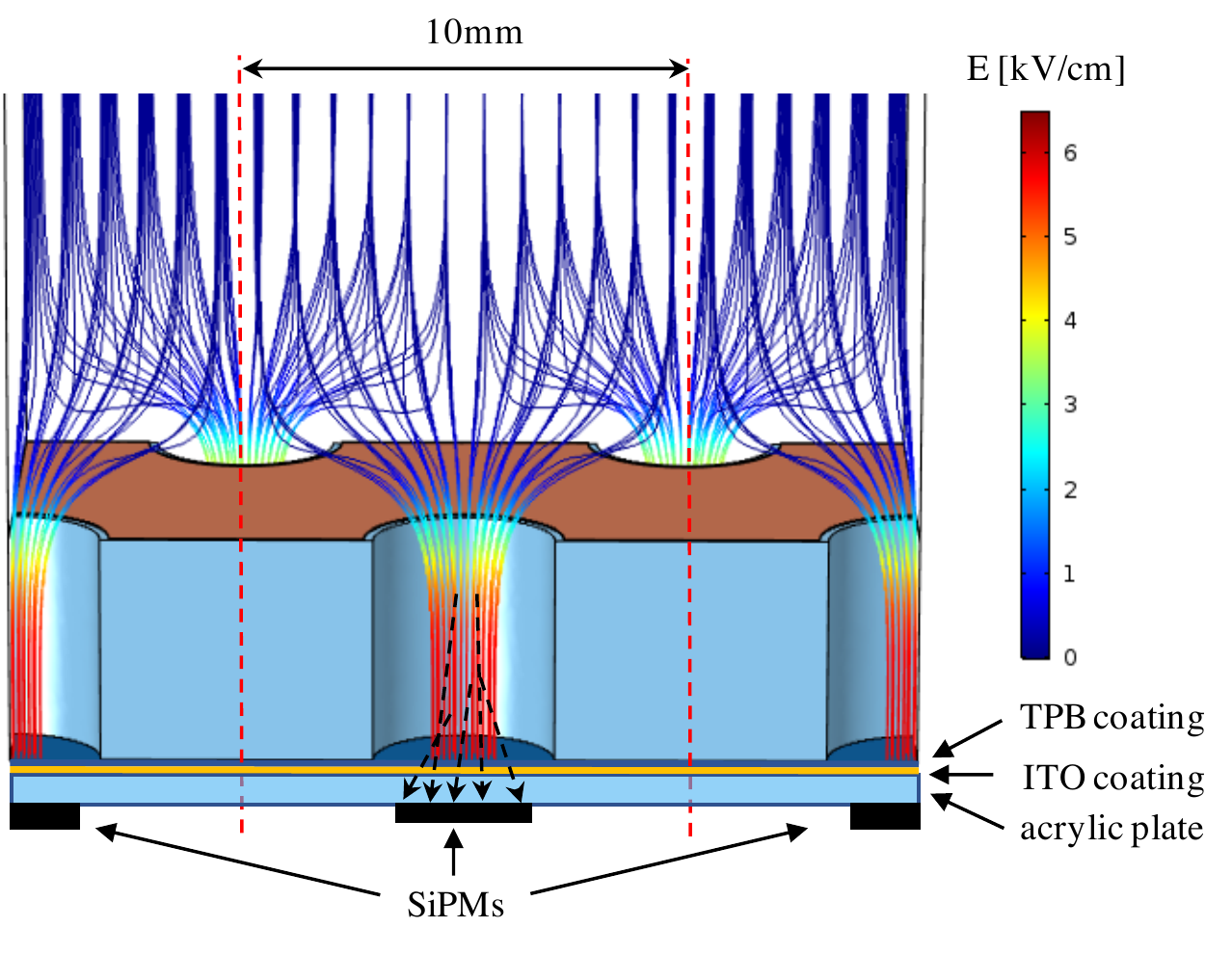}
    \caption{Top: schematic representation of the working principle of the EL-TPC. Photons scattered at the sample reach the xenon gas, creating ionization clusters that drift, while diffusing, towards the anode plane, where they induce electroluminescence. Bottom: close-up of the electroluminescence region, based on the recently introduced acrylic-based electroluminescence multipliers, developed in collaboration between IGFAE and the CERN-RD51 workshops \cite{FATGEM}.}
    \label{fig:sketch}
\end{figure}

The ionization clouds (hereafter `clusters') drift, due to the electric field $E_{drift}$ of the TPC, towards the electroluminescence/anode plane, as shown in Fig.~\ref{fig:sketch}-top, following a diffusion law as a function of the drift distance $z$:
\begin{equation}
\sigma_{z (x,y)} = D_{L (T)}^* \sqrt{z} \label{diffusion_law}
\end{equation}
where $D_L^*$ and $D_T^*$ are the longitudinal and transverse diffusion coefficients, respectively. In fact, diffusion is impractically large in pure noble gases, given that the cooling of ionization electrons is inefficient under elastic collisions only. Addition of molecular additives, enabling vibrational degrees of freedom at typical electron energies, is a well established procedure known to improve the situation drastically, and can be accurately simulated with the electron transport codes Magboltz/Pyboltz \cite{Magboltz, Pyboltz}. In particular, a small (0.4\%) addition of CH$_4$ is sufficient to reduce the cluster size well below that in pure xenon (Fig. \ref{fig:ClusterSize}-bottom), as required for photon-counting. An essential ingredient to the use of Xe-CH$_4$ admixtures is the recent demonstration that the electroluminescence signal is still copious in these conditions \cite{Henriques}.\footnote{This unanticipated result, that might not look significant at first glance, results from a very subtle balance between the quenching of the xenon triplet state and the cooling of drifting electrons through inelastic collisions \cite{DiegoMicro}.} Hence, for a drift field $E_{drift} = 110$~V/cm, the cluster's longitudinal size can be kept at the $\sigma_z = 4$~mm level even for a 50~cm-long drift, corresponding to a temporal spread of $\sigma_t = 0.75~{\mu}$s, while the transverse size approaches $\sigma_{x,y} = $10~mm. The electron drift velocity is $v_d=\sigma_z/\sigma_t=$5 mm/$\mu$s.

The proposed detection concept is depicted in Fig. \ref{fig:sketch}-top, with Fig. \ref{fig:sketch}-bottom displaying a close-up of the pixelated readout region, that relies on the recent developments on large-hole acrylic multipliers \cite{FATGEM}.
Provided sufficient field focusing can be achieved at the structure, as shown in Fig. \ref{fig:sketch}-bottom, the ionization clusters will enter a handful of holes, creating a luminous signal in the corresponding silicon photomultiplier (SiPM) situated right underneath, thus functioning, in effect, as a pixelated readout. 
In summary: i) x-rays that Compton-scatter at the sample interact with the xenon gas and give rise to clusters of characteristic size somewhere in the range 1-10 mm-$\sigma$, depending on the distance to the electroluminescence plane; ii) given the relatively large x-ray mean free path of around 20~cm in xenon at 1~bar, one anticipates a sparse distribution of clusters, that can be conveniently recorded with 10~mm-size pixels/pads, on a readout area of around 2000 cm$^2$ ($N_{pix}=2000$).

From the FWHM per x-ray cluster at about mid-chamber: $\Delta_{x,y}|_{x-ray} = 2.35/\sqrt{2} \cdot \sigma_{x,y} = 16$~mm, an average multiplicity $M$ of around $4$ per cluster may be assumed if resorting to $10~\textnormal{mm} \times 10~\textnormal{mm}$ pixels/pads. The temporal spread, on the other hand, can be approximated by: ${\Delta}T_{x-ray} = 2.35/\sqrt{2} \cdot \sigma_z/v_d = 1.35~{\mu}$s. Taking as a reference an interaction probability of $P_{int} = 2.9\times10^{-4}$ ($5~{\mu}$m water-equivalent cell, 10~mm of air), a 70\% detection efficiency $\epsilon$, and an $m=20$\% pixel occupancy, this configuration yields a plausible estimate of the achievable counting rate as:
\begin{equation}
r_{max} = \frac{1}{\epsilon P_{int}} \frac{m \cdot N_{pix}}{M} \frac{1}{\Delta{T}_{x-ray}} = 3.6 \times 10^{11}~\textnormal{(ph/s)}    
\end{equation}
compatible a priori with the beam rates for hard x-rays foreseen at the new generation of light sources \cite{ESRF}. However, in order to have a realistic estimate of the actual counting performance it is imperative to understand which level of occupancy/pile-up can be \emph{really} tolerated by the detector, before the photon-counting performance deteriorates above the Poisson-limit or proportionality of response is irreparably lost. We address this problem specifically in section \ref{TPCresponse}. 

\subsection{Geometry optimization with Geant4}

\begin{figure}[h!]
    \centering
    \includegraphics[width=80mm]{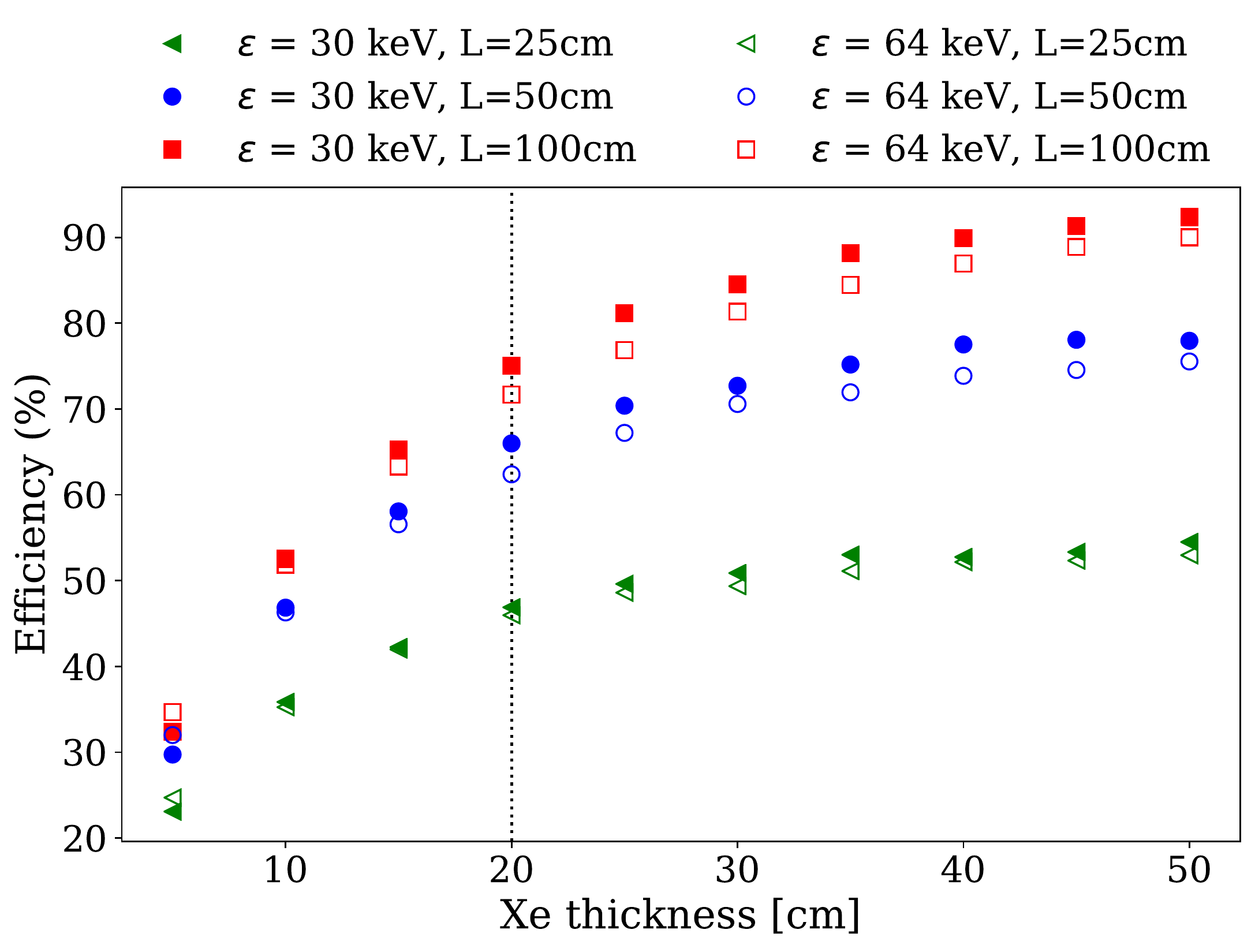}
    \caption{Efficiency as a function of the thickness of the xenon cylinder ($R_o$-$R_i$) for different lengths, at energies of 30 and 64~keV. The dotted line indicates the benchmark geometry considered in text, for a length $L=50$~cm.}
    \label{fig:efficiency}
\end{figure}

The suitability of the TPC technology for SCXM depends primarily on the ability to detect $\sim60$ keV photons within a realistic gas volume, in the absence of pressurization. Given that the mean free path of 60~keV x-rays in xenon is 20~cm, the most natural $4\pi$-geometry adapting to this case is a hollow cylinder with a characteristic scale of around half a meter. On the other hand, the geometrical acceptance is a function of $\arctan(2R_{i}/L)$, with $L$ being the length and $R_{i}$ the inner radius of the cylinder. In order to place the sample holder, step motor, pipes and associated mechanics, we leave an $R_i= 5$~cm inner bore. Finally, the xenon thickness ($R_o$-$R_i$), that is the difference between the outer and inner TPC radii, becomes the main factor for the detector efficiency, as shown in Fig.~\ref{fig:efficiency}. We discuss two photon energies: 30 and 64~keV. The latter represents the theoretical optimum for SCXM in terms of dose, while the former, sitting just below the $K$-shell energy of xenon, is a priory more convenient for counting due to the absence of characteristic (K-shell) x-ray re-emission inside the chamber. The mean free path is similar for the two energies, therefore no obvious advantage (or disadvantage) can be appreciated in terms of detector efficiency, at this level of realism.

\begin{figure}[h!]
    \centering
    \includegraphics[width=85mm]{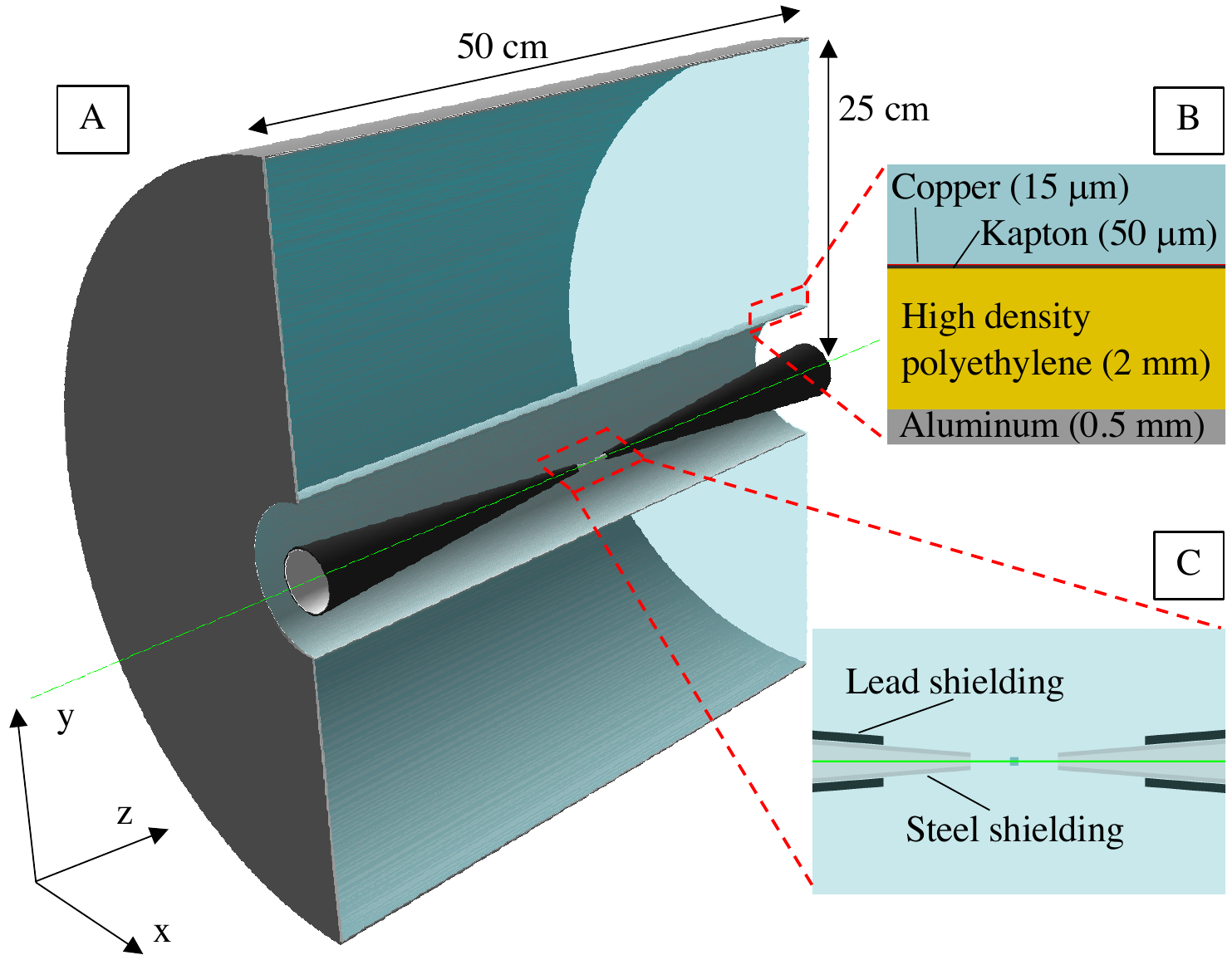}
    \caption{A) TPC geometry in Geant4, aimed at providing nearly $4\pi$-coverage for SCXM. B) detail of the region faced by x-rays when entering the detector, that includes the vessel and field cage. C) detail of the sample region and the shielding cones.}
    \label{fig:TPC3D}
\end{figure}

We consider now a realistic geometry, opting for an inner cylinder shell made out of 0.5~mm-thick aluminum walls, with 2~mm HDPE (high density polyethylene), $50~{\mu}$m kapton and $15~{\mu}$m copper, sufficient for making the field cage of the chamber, that is needed to minimize fringe fields (inset in Fig. \ref{fig:TPC3D}). The HDPE cylinder can be custom-made and the kapton-copper laminates are commercially available and can be adhered to it by thermal bonding or epoxied, for instance. The external cylinder shell may well have a different design, but it has been kept symmetric for simplicity. We consider in the following a configuration that enables a good compromise in terms of size and flexibility: $L=50~$cm and $R_o= 25~$cm. The geometrical acceptance nears in this case 80\%. Additional 10~cm would be typically needed, axially, for instrumenting the readout plane and taking the signal cables out of the chamber, and another 10~cm on the cathode side, for providing sufficient isolation with respect to the vessel, given that the voltage difference will near 10~kV. Although those regions are not discussed here in detail, and have been replaced by simple covers, the reader is referred to \cite{Francesc} for possible arrangements. With these choices, the vessel geometry considered in simulations is shown in Fig. \ref{fig:TPC3D}, having a weight below 10~kg.

The necessary structural material of the walls and the presence of air in the hall reduce the overall efficiency from 62.8\% to 58.5\% (64~keV) and from 64.5\% to 40.0\% (30~keV). The beam enters the experimental setup from the vacuum pipes (not included in the figure) into two shielding cones (made of stainless steel and covered with lead shields) and from there into the sample region. Our case study is that of a 33~nm DNA feature inside a $5~{\mu}$m cell, and 5~mm air to and from the shielding cones. The conical geometry is conceived not to crop the angular acceptance of the x-rays scattered on-sample, providing enough space to the focusing beam, and enabling sufficient absorption of stray x-rays from beam-air interactions along the pipes. In a $4\pi$ geometry as the one proposed here, the cell holder and step motor should ideally be placed along the polarization axis, where the photon flux is negligible.

\subsection{Image formation in the TPC}

The parameters used for computing the TPC response rely largely on the experience accumulated during the NEXT R\&D program. We consider a voltage of -8.5~kV at the cathode and 3~kV across the electroluminescence structure, with the anode sitting at ground, a situation that corresponds to fields around $E_{drift}=110$~V/cm and $E_{el}=6$~kV/cm in the drift and electroluminescence regions, respectively. The gas consists of Xe/CH$_4$ admixed at 0.4\% in volume in order to achieve a 40-fold reduction in cluster size compared to operation in pure xenon (Fig. \ref{fig:ClusterSize}-bottom). The electroluminescence plane will be optically coupled to a SiPM matrix, at the same pitch, forming a pixelated readout. The optical coupling may be typically done with the help of a layer of ITO (indium-tin oxide) and TPB (tetraphenyl butadiene) deposited on an acrylic plate, following \cite{Francesc}. This ensures wavelength shifting to the visible band, where SiPMs are usually more sensitive. The number of SiPM-photoelectrons per incoming ionization electron, $n_{phe}$, that is the single most important figure of merit for an EL-TPC, can be computed from the layout in Fig. \ref{fig:sketch}-bottom, after considering: an optical yield $Y = 250$ ph/e/cm at $E_{el}=6$~kV/cm \cite{FATGEM}, a TPB wavelength-shifting efficiency $WLSE_{TPB}=0.4$ \cite{Gehman}, a solid angle coverage at the SiPM plane of $\Omega_{SiPM}=0.3$ and a SiPM quantum efficiency $QE_{SiPM}=0.4$. Finally, according to measurements in \cite{Henriques}, the presence of 0.4\% CH$_4$ reduces the scintillation probability by $P_{scin}=0.5$, giving, for a $h=5$~mm-thick structure:
\begin{equation}
n_{phe} = Y \cdot h \cdot WLSE_{TPB} \cdot \Omega_{SiPM} \cdot QE_{SiPM} \cdot P_{scin} = 3   
\end{equation}
 
Since the energy needed to create an electron-ion pair in xenon is $W_I=22$~eV, each 30-64~keV x-ray interaction will give raise to a luminous signal worth 4000-9000 photoelectrons (phe), spanning over 4-8 pixels, hence well above the SiPM noise. The energy resolution (FWHM) is obtained from \cite{Henriques} as:
\begin{equation}
\mathcal{R}(\varepsilon\!=\!64~\textnormal{keV}) \simeq 2.355 \sqrt{F + \frac{1}{n_{phe}}\left(1+\frac{\sigma_G^2}{G^2}\right)}\sqrt{\frac{W_I}{\varepsilon}} = 3.1\%
\end{equation}
with $\sigma_G/G$ being the width of the single-photon distribution (around 0.1 for a typical SiPM) and $F\simeq0.17$ the Fano factor of xenon. For comparison, a value compatible with $\mathcal{R}(\varepsilon\!=\!64~\textnormal{keV})=5.5\%$ was measured for acrylic-hole multipliers in \cite{FATGEM}. In present simulations, the contribution of the energy resolution has been included as a gaussian smearing in the TPC response.

Finally, the time response function of the SiPM is included as a Gaussian with a 7~ns width, convoluted with the transit time of the electrons through the electroluminescence structure $\Delta{T}_{EL} = 0.36~\mu{s}$, being both much smaller in any case than the typical temporal spread of the clusters (dominated by diffusion). The sampling time is taken to be ${\Delta}T_s = 0.5~\mu$s as in \cite{Francesc}, and a matrix of 1800 10~mm-pitch SiPMs is assumed for the readout. Images are formed after applying a 10~phe-threshold to all SiPMs.

\begin{figure}[h]
    \centering
    \includegraphics[width=80mm]{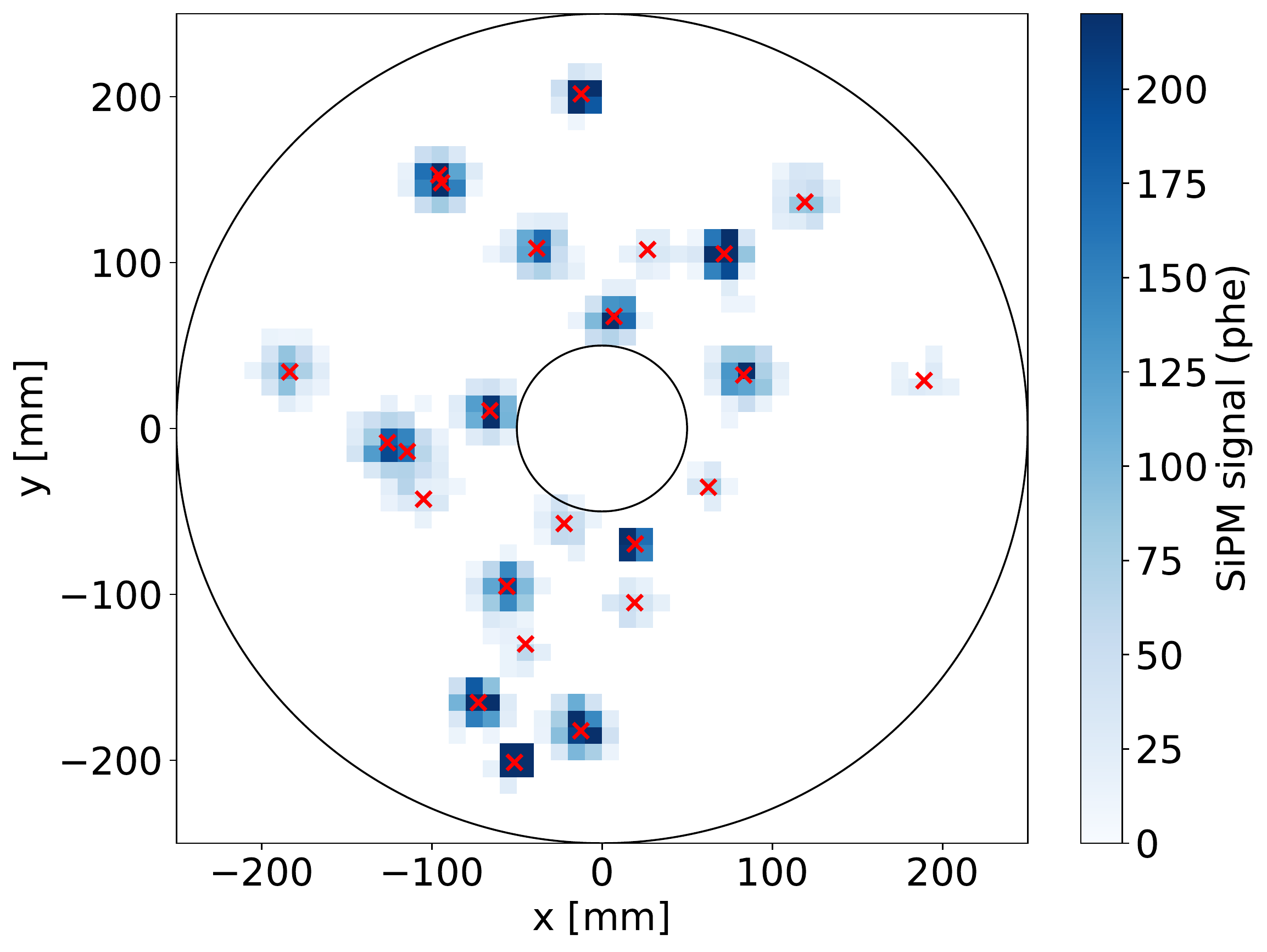}
    \caption{A typical TPC image reconstructed from the SiPM signals (in phe), as recorded in one time-slice (${\Delta}T_s=~0.5~ {\mu}$s), for a beam rate of $r=3.7\times 10^{10} ~s^{-1}$. The crosses show the clusters' centroids, obtained from `MC-truth' information.}
    \label{fig:ClusterCounting}
\end{figure}

A fully processed TPC image for one time slice (${\Delta}T_s=~0.5~ {\mu}$s), obtained at a beam rate of $r=3.7\times 10^{10}$~ph/s for a photon energy $\varepsilon=64$~keV, is shown in Fig. \ref{fig:ClusterCounting}. The main clusters have been marked with crosses, by resorting to `Monte Carlo truth', i.e., they represent the barycenter of each primary ionization cluster in Geant4. The beam has been assumed to be continuous, polarized along the $x$-axis, impinging on a 5~${\mu}$m water cube surrounded by air, with a 33~nm DNA cubic feature placed at its center. The Geant4 simulations are performed at fixed time, and the x-ray interaction times are subsequently distributed uniformly within the dwell time corresponding to each position of the scan. It must be noted that interactions taking place at about the same time may be recorded at different times depending on the $z$-position of each interaction, (and viceversa, clusters originating at different interaction times, may eventually be reconstructed in the same time slice). This scrambling (unusual under typical TPC operation) renders every time slice equivalent for the purpose of counting. In principle, the absolute time and $z$ position can be disambiguated from the size of the cluster, using the diffusion relation in eq. \ref{diffusion_law}, thus allowing photon-by-photon reconstruction in time, space, and energy. A demonstration of the strong correlation between $z$-position and cluster width, for 30~keV x-ray interactions, can be found in \cite{DiegoAccurate} for instance.

The design parameters used in this subsection are compiled in tables 1-4 of the Appendix~\ref{appendixB}.

\section{Photon counting capabilities}\label{TPCresponse}

\subsection{Ideal counting limit}

The attenuation in the structural materials, re-scatters, characteristic emission, as well as the detector inefficiency, are unavoidable limiting factors for counting. These intrinsic limitations can be conveniently evaluated from the signal-to-noise ratio, defined from the relative spread in the number of ionization clusters per scan step (see Fig. \ref{fig:Dose}), as obtained in Monte Carlo ($n_{MC}$): 
\begin{equation}
S/N = n_{MC}/\sigma_{n_{MC}} \label{S/N1}
\end{equation}

\begin{figure}[h]
    \centering
    \includegraphics[width=85mm]{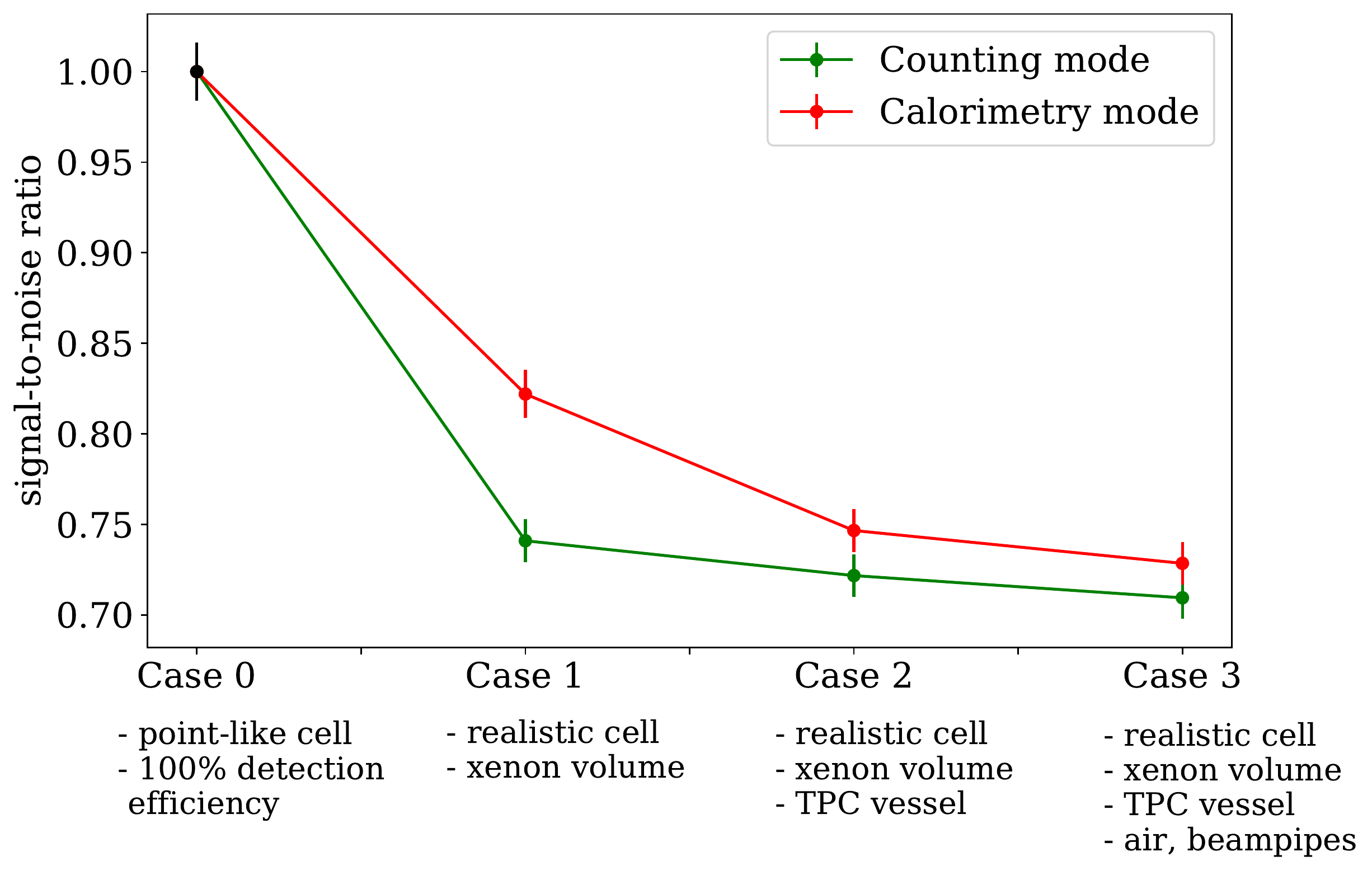}
    \caption{Intrinsic counting performance (using Monte Carlo truth information) for 64~keV x-ray photons, characterized by the signal to noise ratio (relative to case 0). Photon counting (green) and calorimetric mode (red) are displayed as a function of the realism of the simulations.}
    \label{fig:simu_realism}
\end{figure}

Figure \ref{fig:simu_realism} shows the deterioration of the $S/N$ for 64~keV photons, as the realism of the detector increases. It has been normalized to the relative spread in the number of photons scattered on-sample per scan step, $\sqrt{N_0}$, so that it equals 1 for a perfect detector (see appendix~\ref{appendixA}):
\begin{equation}
S/N^* \equiv \frac{1}{\sqrt{N_0}} \cdot S/N \label{S/N2}
\end{equation}

The figure also shows the $S/N^*$ in `calorimetric mode', with the counting performed by simply integrating the total collected light per scan step ($\varepsilon_{tot}$), instead of photon-by-photon. $S/N^*$ is defined in that case, equivalently, as: $S/N^* = (\varepsilon_{tot}/\sigma_{\varepsilon_{tot}}) / \sqrt{N_0}$. The values obtained are just slightly below the ones expected considering detector inefficiency alone (see appendix~\ref{appendixA}):
\begin{equation}
S/N^* \simeq \sqrt{\epsilon} 
\end{equation}
therefore suggesting a small contribution from re-scatters in the materials or other secondary processes.

\subsection{Real counting}

\begin{figure}[h]
    \centering
    \includegraphics[width=85mm]{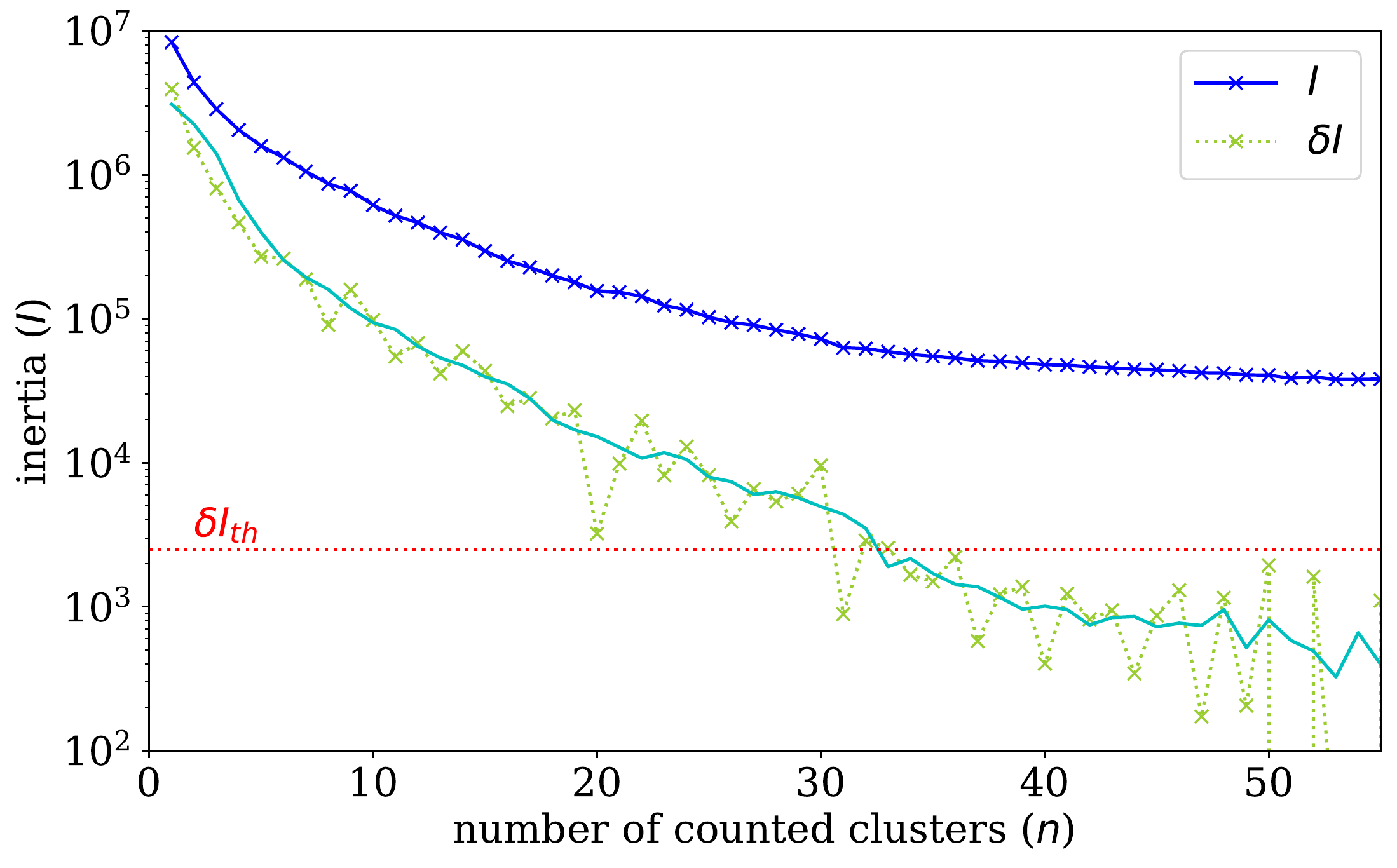}
        \caption{The K-means cluster-counting algorithm evaluates the partition of $N$ observations (voxelized ionization clusters in our case) in $n$ clusters, so as to minimize the inertia $I$, defined as the sum of the squared distances of the observations to their closest cluster center. In the plot: convergence of K-means for a beam rate of 10$^{11}$ ph/s. A Savitzky–Golay filter is applied for the purpose of smoothing the variation of the inertia $\delta I$.}
    \label{fig:kmeans}
\end{figure}

\begin{figure}[h]
    \centering
    \includegraphics[width=85mm]{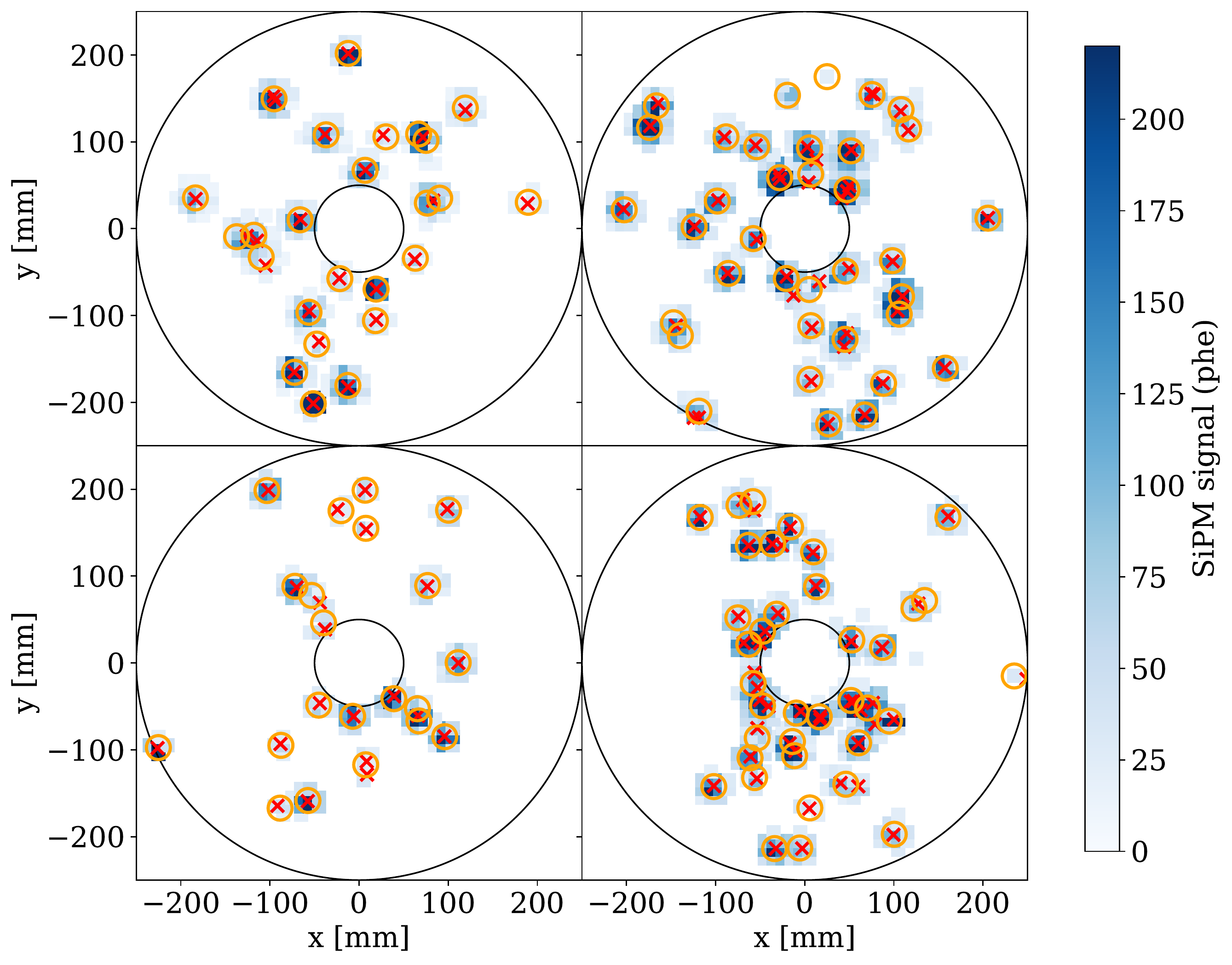}
        \caption{Cluster counting performance for typical ${\Delta}T_s=~0.5~{\mu}$s time-slices, for different energies ($\varepsilon$) and beam rates ($r$). Crosses indicate the cluster centroids from MC and circles are the clusters found by K-means. The average counting-efficiency and purity along the detector are given below in brackets. Top left: $\varepsilon$ = 64~keV and $r = 3.7\times10^{10}$~ph/s ($\epsilon_{counting}$ = 88.2\%, $p_{counting}$ = 86.9\%). Top right: $\varepsilon$ = 64~keV and $r = 7.5\times10^{10}$~ph/s ($\epsilon_{counting}$ = 84.2\%, $p_{counting}$ = 83.2\%). Bottom left: $\varepsilon$ = 30~keV and $r = 6.5\times10^{10}$~ph/s ($\epsilon_{counting}$ = 87.9\%, $p_{counting}$ = 87.5\%). Bottom right: $\varepsilon$ = 30~keV and $r = 1.3\times10^{11}$~ph/s ($\epsilon_{counting}$ = 83.9\%, $p_{counting}$ = 83.1\%). For $\varepsilon$ = 30~keV only about half of the clusters are produced, which enables measuring at higher beam rates than $\varepsilon$ = 64~keV, at comparable efficiency and purity.}
    \label{fig:counting_beam_rate1}
\end{figure}

Given the nature of the detector data (Fig. \ref{fig:ClusterCounting}), consisting of voxelized ionization clusters grouped forming ellipsoidal shapes, generally separable, and of similar size, we select the K-means clustering method~\cite{Kmeans} to perform cluster counting. The counting algorithm has been implemented as follows:
i) the `countable' clusters are first identified time-slice by time-slice using Monte Carlo truth information, as those producing a signal above a certain energy threshold ($\varepsilon_{th}$) in that slice. The energy threshold is chosen to be much lower than the typical cluster energies. In this manner, only small clusters are left out of the counting process when most of their energy is collected in adjacent time-slices from which charge has spread out due to diffusion, and where they will be properly counted once the algorithm is applied there; ii) a weighted inertia ($I$) distribution is formed, as conventionally done in K-means, and a threshold ($\delta I_{th}$) is set to the variation of the inertia with the number of clusters counted by the algorithm ($n$) (Fig. \ref{fig:kmeans}). The threshold is optimized for each beam rate condition. We concentrate on beam rates for which the average efficiency and purity of the cluster identification in 2d slides is larger than 80\%, as the ones illustratively depicted in Fig. \ref{fig:counting_beam_rate1}. The counting efficiency and purity can been defined, as customary, as:
\begin{eqnarray}
\epsilon_{counting} & = & \frac{n_{matched}}{n_{MC}} \label{eff_count}\\
p_{counting}        & = & \frac{n_{matched}}{n}
\end{eqnarray}
where $n_{matched}$ is the number of counted clusters correctly assigned to MC clusters and $n_{MC}$ is the number of MC clusters. The K-means optimization parameters have been chosen to simultaneously maximize the counting efficiency while achieving $n \simeq n_{MC}$, therefore $\epsilon_{counting} \simeq p_{counting}$.

\begin{figure}[h]
    \centering
    \includegraphics[width=85mm]{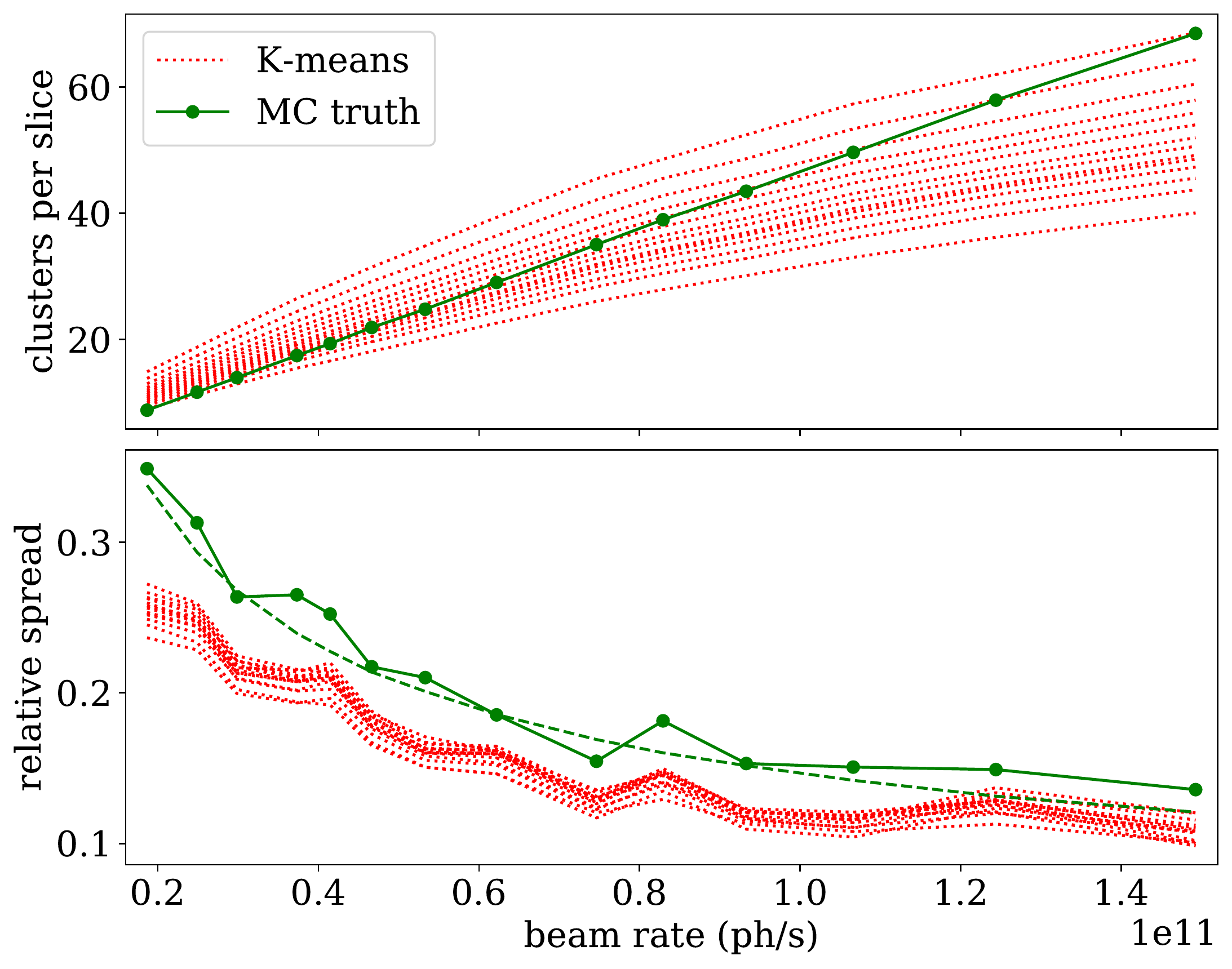}
        \caption{Top: counting performance characterized through the average number of clusters counted per 2d time-slice as a function of the beam rate for $\varepsilon$~=~64~keV. Bottom: relative spread of the number of clusters per 2d time-slice from Monte Carlo truth and counted with K-means. The $1/\sqrt{r}$ expectation (dashed) is shown for comparison.}
    \label{fig:counting_beam_rate2}
\end{figure}

Fig. \ref{fig:counting_beam_rate2} (top) shows the performance of the counting algorithm, presenting the average number of clusters counted per 2d slice as a function of beam rate, with $\varepsilon_{th}$ and $\delta I_{th}$ optimized for each case as described above (green line). Red lines indicate the predictions outside the optimized case, that illustrate the consistent loss of linearity as the beam rate increases. Fig. \ref{fig:counting_beam_rate2} (bottom) shows the relative spread in the number of counted clusters $\sigma_n/n$, and comparison with Monte Carlo truth. These results can be qualitatively understood if recalling that, by construction, the threshold inertia is strongly correlated with the average number of clusters and its size. Therefore, a simple K-means algorithm will inevitably bias the number of counted clusters to match its expectation on $I$, if no further considerations are made. Therefore, once $\delta I_{th}$ has been adjusted to a certain beam rate, there will be systematic overcounting for lower beam rates, and undercounting for higher ones, as reflected by Fig. \ref{fig:counting_beam_rate2} (top). In present conditions, a 2$^{\scriptsize{\textnormal{nd}}}$ order polynomial is sufficient to capture this departure from proportionality introduced by the algorithm. A similar (although subtler) effect takes place for the cluster distributions obtained slice-by-slice, where this systematic overcounting-undercounting effect makes the cluster distribution marginally (although systematically) narrower, as seen in Fig. \ref{fig:counting_beam_rate2} (bottom). As a consequence, the directly related magnitude $S/N^*$ (eqs. \ref{S/N1}, \ref{S/N2}), is not deteriorated by the counting algorithm. On the other hand, proportionality is lost, and its impact needs to be addressed, depending on the application. The particular case of SCXM is scrutinized in the next section.

Finally, the photon-counting efficiency (eq. \ref{eff_count}) can be assessed through Fig.~\ref{fig:counting_efficiency}-top, where it is displayed as a function of the beam rate on target. It can be seen how, for the case of 30 and 64~keV photons, its value exceeds 85\% for rates up to 10$^{11}$~ph/s and $0.5\cdot10^{11}$~ph/s, respectively. At these high beam rates, counting capability suffers from event pile-up while, at low beam rates, it is limited by the presence of low-energy deposits (corresponding to x-ray interactions for which most of the energy is collected in adjacent slices). It must be recalled, at this point, that a complete reconstruction requires combining 2d time-slices as the ones studied here, in order to unambiguously identify clusters in 3d. Given that each cluster extends over 4-6 slices due to diffusion, and clusters are highly uncorrelated, a 3d counting efficiency well above 90\% can be anticipated in the above conditions.

\section{Projections for SCXM}\label{results}
We propose the characterization of the EL-TPC technology in light of its performance as a cellular microscope, through the study of the smallest resolvable DNA-feature (size $d$) as a function of the scan time ($\Delta{T}_{scan}$). Justification of the following derivations can be found in appendix~\ref{appendixA}, starting with:
\begin{equation}
d = \left(R^2 2 l^2\frac{(l\lambda_w^{-1} + 2a\lambda_a^{-1})}{(\lambda_f^{-1} - \lambda_w^{-1})^2} \frac{1}{C_l(r)^2 \cdot S/N^{*,2}\cdot r \cdot\Delta{T}_{scan}}\right)^{1/4} \label{eq:Clin}    
\end{equation}
Here $R$ equals 5 under the Rose criterion and the rate-dependent coefficient $C_l<1$ depends on the deviation of the counting algorithm from the proportional response, its expression being given in appendix~\ref{appendixA}. Other magnitudes have been already defined. Since the smallest resolvable feature size ($d^{\dagger}$) is ultimately determined by the dose imparted at it when structural damage arises (eq. \ref{surDos}, Fig. \ref{fig:Dose}), the necessary scan time to achieve such performance ($\Delta{T}_{scan}^{\dagger}$) can be readily obtained:
\begin{equation}
\Delta{T}_{scan}^{\dagger} = R^2 2 l^2\frac{(l\lambda_w^{-1} + 2a\lambda_a^{-1})}{(\lambda_f^{-1} - \lambda_w^{-1})^2} \frac{1}{C_l(r)^2 \cdot S/N^{*,2}\cdot r \cdot (d^{\dagger})^4} \label{eq:DeltaTdagger}    
\end{equation}
For a detector with finite efficiency, the value of $d^{\dagger}$ can be recalculated by simply accounting for the necessary increase in fluence (and hence in dose), as:
\begin{eqnarray}
\phi \rightarrow \phi' & = & \phi/\epsilon \\
\mathcal{D} \rightarrow \mathcal{D}' & = & \mathcal{D}/\epsilon
\end{eqnarray}
that results in slightly deteriorated values compared to Fig. \ref{fig:Dose}: $d^{\dagger}=36$~nm instead of $d^{\dagger}=33$~nm for $\varepsilon$=64~keV, and $d^{\dagger}=44$~nm instead of $d^{\dagger}=37$~nm for $\varepsilon$=30~keV.

\begin{figure}[h]
    \centering
    \includegraphics[width=85mm]{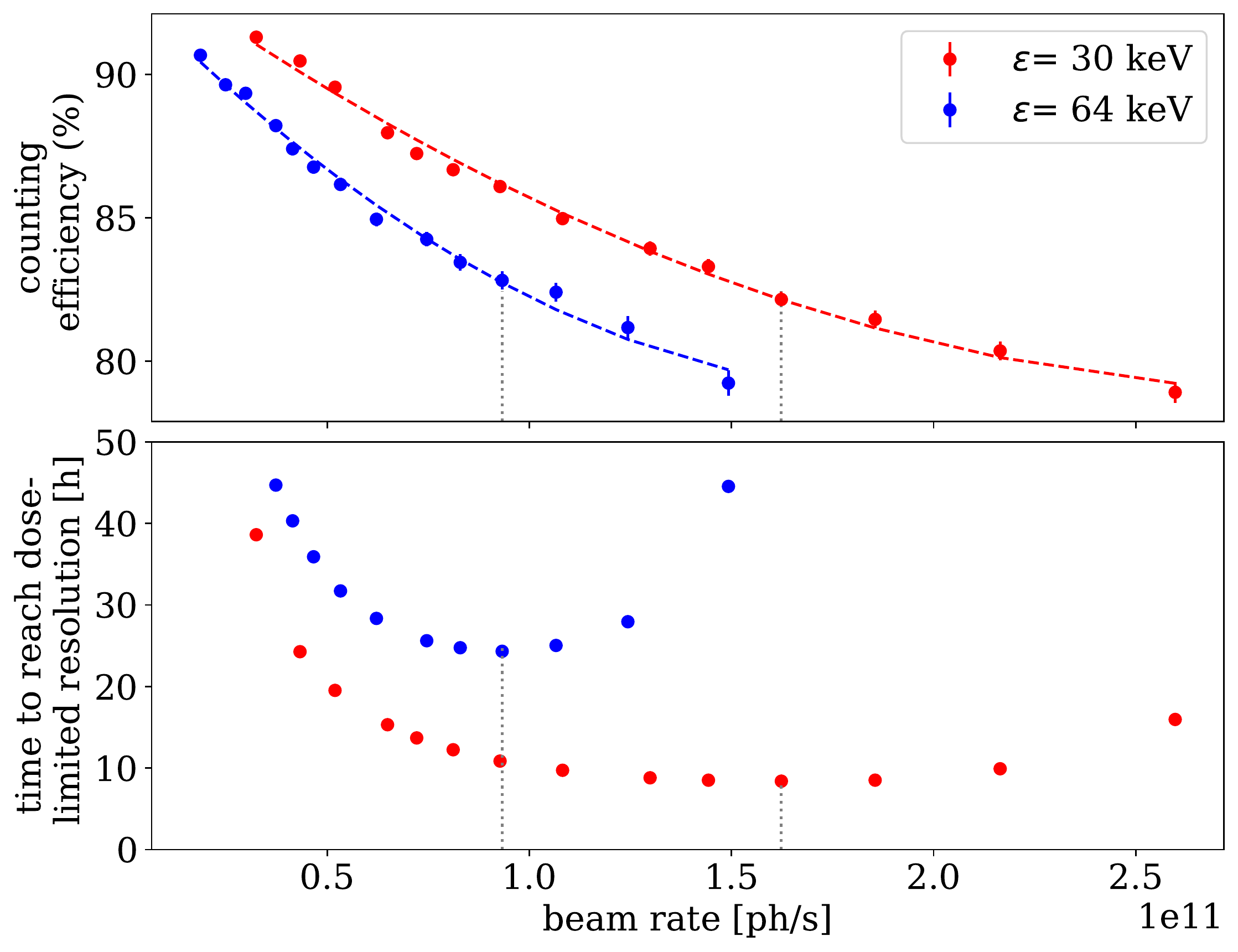}
        \caption{Top: efficiency of the cluster counting process as a function of the beam rate for x-rays of 30 and 64~keV. Bottom: time to reach the dose-limited resolution as a function of the beam rate. A minimum is reached when the product of $C_l^2\cdot r$ reaches a maximum, i.e. the time decreases with beam rate until the effect of the non-proportional counting (resulting from event pile-up) becomes dominant. The optimum beam rate and corresponding counting efficiency are marked with a dotted line for both energies.}
    \label{fig:counting_efficiency}
\end{figure}

The limiting scan time (i.e., above which structural damage will appear) can be hence assessed from the behaviour of eq. \ref{eq:DeltaTdagger} with beam rate, as shown in Fig.~\ref{fig:counting_efficiency}-bottom. For 64~keV, the loss of linearity of the counting algorithm at high rates results in a turning point at $9.3 \times 10^{10}$ ph/s, above which an increase in rate stops improving the ability to resolve an image. For 30~keV, due to the absence of characteristic emission, only about half of the clusters are produced and the optimum rate is found at a higher value, $r = 1.6 \times 10^{11}$. The counting efficiency and purity in these conditions is in the range 82-84\%.

\begin{figure}[h]
    \centering
    \includegraphics[width=85mm]{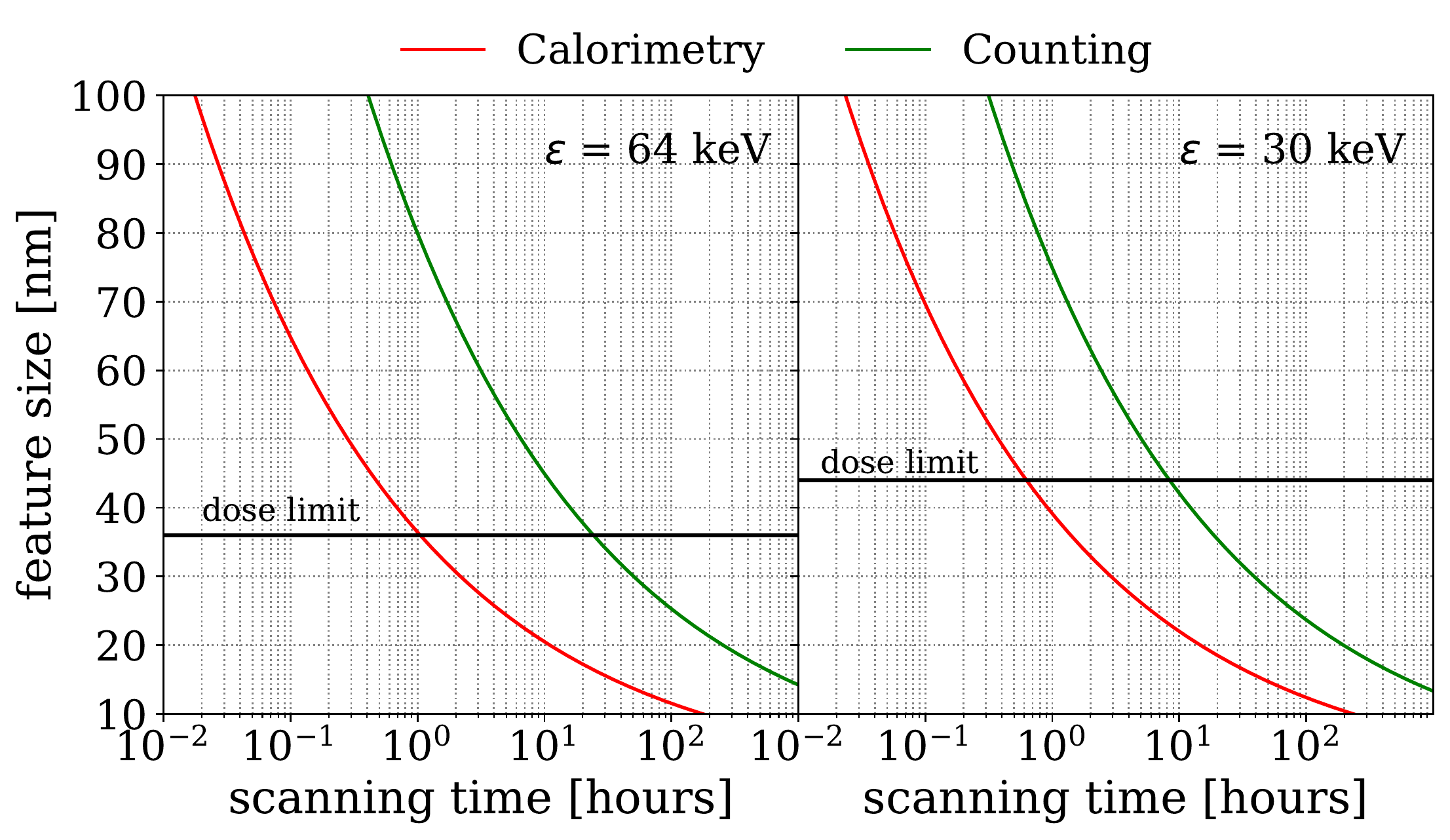}
    \caption{Resolution achievable with a 64~keV photon beam (left) and a 30~keV photon beam (right) as a function of the scan time for a cell of $5$~$\mu$m (green line). The red line shows the limit in which a calorimetric measurement is performed and photon-by-photon counting is abandoned. The horizontal line shows the dose-limited resolution in each case, prior to inducing structural damage.}
    \label{fig:ScanTime}
\end{figure}

It is now possible to evaluate eq. \ref{eq:Clin} under different scenarios: i) a relatively simple calorimetric mode (total energy is integrated), for which we assume a hard x-ray beam rate typical of the new generation of synchrotron light sources as $r = 10^{12}$~ph/s, and ii) a rate-limited photon-by-photon counting scenario, for the optimum rates $r = 9.3\times10^{10}$ ph/s (64 keV) and $r=1.6\times10^{11}$ ph/s (30 keV), obtained above. Values for $C_l(r)$ are extracted from 2$^{\scriptsize\textnormal{nd}}$-order fits as discussed in appendix. The remaining parameters are common to both modes: $S/N^*=0.71$, efficiency $\epsilon=58.5\%$  (64~keV), $S/N^*=0.63$, $\epsilon=40.0\%$ (30~keV); finally we assume $l=5~{\mu}$m, $a=5$ mm, $R=5$, with the mean free paths ($\lambda$) taken from table \ref{tab:material_parameters}. Results are summarized in Fig. \ref{fig:ScanTime}. At 64~keV, the dose-limited resolution $d^{\dagger}=36$~nm can be achieved in approximately 24~h while, at $30$~keV, $d^{\dagger}=44$~nm is reached in just 8~h. In the absence of systematic effects, operation in calorimetric mode would bring the scan time down to $\leq 1$~h in both cases, although abandoning any photon-by-photon counting capabilities.

\section{Discussion}\label{discussion}

The results presented here illustrate the potential of the proposed technology for high energy x-ray detection (up to $\simeq 60$-$70$ keV) at high-brightness synchrotron light sources, in particular for cellular imaging. In deriving them, we have adopted some simplifications, that should be superseded in future work, and are analyzed here:
\begin{enumerate}
    \item \emph{Availability of photon-by-photon information}: cluster reconstruction with high efficiency and purity enables $x, y, t + t_{drift}$ and $\varepsilon$ determination, and arguably the interaction time $t$ and $z$ position can be obtained from the study of the cluster size, as it has been demonstrated for 30~keV x-rays at near-atmospheric pressure before \cite{DiegoAccurate}. This can help at removing backgrounds not accounted for, as well as any undesired systematic effect (beam or detector related). Since this technique provides a parallax-free measurement, the concept may be extended to other applications, e.g., x-ray crystallography. The presence of characteristic emission from xenon will unavoidably create confusion, so if unambiguous correspondence between the ionization cluster and the parent x-ray is needed, one must consider operation at $\lesssim$ 30~keV.
    
    \item \emph{Data processing and realism}: photon-by-photon counting at a rate nearing $5\cdot10^7$~ph/s over the detector ($\equiv$ 10$^{11}$~ph/s over the sample) , as proposed here, is a computer intensive task. Achieving this with sufficient speed and accuracy will require the optimization of the counting algorithm, something that will need to be accomplished, ultimately, with real data. To this aim, both the availability of parallel processing as well as the possibility of simultaneous operation in calorimetric mode are desirable features. This will be studied in the near future through a dedicated experiment.
    
    \item \emph{Simplicity and compactness}: the detector geometry proposed here has been conceived as a multi-purpose permanent station. A portable device focused purely on SCXM, on the other hand, could simply consist of a cubic $25\textnormal{cm} \times 25\textnormal{cm} \times 25\textnormal{cm}$ vessel that may be positioned, e.g., on top of the sample (at a distance of about $\sim 5$cm). The geometry would thus have an overall efficiency around 30\% for 64~keV photons. For SCXM, and given that $S/N^* \simeq \sqrt{\epsilon}$ as shown in this work, a loss of efficiency can be almost fully compensated by means of the corresponding increase in beam rate, at the price of a deteriorated value for the dose limited resolution $d^{\dagger}$. In this case, a value corresponding to $d^{\dagger}=41$~nm could be achieved in 12~h, for our test study.
    
    \item \emph{Feasibility}: the technology proposed comes from the realm of high energy physics, with an inherent operational complexity that might not be affordable at light source facilities. A further possibility could be considered, by resorting to ultra-fast (1.6~ns resolution) hit-based TimePix cameras (e.g., \cite{TimePix, Nom19}) with suitable VUV-optics, allowing $256 \times 256$ pixel readout at 80~MHit/s, and thus abandoning completely the SiPM readout. The vessel would just house, in such a case, the acrylic hole multiplier and cathode mesh, together with the power leads; it would be filled with the xenon mixture at atmospheric pressure and interfaced to the outside with a VUV-grade viewport. This would compromise partly the ability to disentangle clusters by using time information, as well as energy information, since only the time over threshold would be stored and not the temporal shape of each cluster, or its energy. On the other hand, it would enhance the spatial information by a factor of $30$ relative to the SiPM matrix proposed here (the hole pitch of the acrylic hole multiplier should be reduced accordingly). Indeed, TimePix cameras are regularly used nowadays for photon and ion counting applications \cite{TimePix1, TimePix2}, but have not been applied to x-ray counting yet, to the best of our knowledge. The counting and signal processing algorithms could be in this way directly ported, given the similarity with the images taken in those applications. The readiness of such an approach, aiming at immediate implementation, represents an attractive and compelling avenue.
\end{enumerate}

The imaging criterion and study case chosen in this work are inspired by \cite{Vil18}, where a dose-limited resolution of 34~nm was obtained for SCXM, compared to around 75~nm for CDI. A typical bio-molecule feature was chosen, embedded in a $5~{\mu}$m cell placed in vacuum. The present study shows that a 36~nm DNA feature can be resolved in similar conditions even after accounting for the presence of beam-shielding, air, photon transport through a realistic detector, including the detector response in detail, and finally implementing photon-counting through a K-means algorithm.

\section{Conclusions and outlook}\label{conclus}
We introduce a new $4\pi$-technology (EL-TPC) designed for detecting $\sim\!60$~keV x-ray photons at rates up to $5\cdot10^7$ ph/s over the detector ($10^{11}$ ph/s over the sample), with an overall detection efficiency (including geometrical acceptance) around 60\%. At these rates, photon-by-photon counting can be achieved at an efficiency and purity above 80\%, and plausibly well above 90\% after straightforward improvements on the counting algorithm employed in this work. The technology has been re-purposed from its original goal in particle physics (the experimental measurement of $\beta\beta0\nu$ decay) and, with a number of minor simplifications, it has been optimally adapted to the task of Compton x-ray microscopy in upcoming light sources. The proposed detector can be implemented either as a permanent station or a portable device. Concentrating on $5~{\mu}$m cells as our test case, we estimate that, under a Rose imaging criterion, and assuming the dose fractionation theorem, 36~nm DNA features may be resolved in 24~h by using a permanent station and 41~nm in 12~h with a portable device.
Alternatively, the scan time could be brought down to less than 1~h by resorting to the calorimetric mode, although the photon-by-photon counting capability would need to be abandoned.
Our analysis includes detailed Geant4 transport, a realistic detector response and a simplified 2d-counting algorithm based on K-means. Thus, we understand that the obtained rate capability (and scan time) should be understood as lower (upper) limits to the actual capabilities when using more refined 3d-algorithms, including constraints in energy and cluster size.

Although substantially below the nominal photon-counting capabilities of solid-state pixelated detectors, we believe that a number of applications could benefit from the proposed development, targeting specifically at the newly available 4$^{\scriptsize\textnormal{th}}$ generation synchrotron light sources, capable of providing high-brightness hard x-rays. Indeed, previous conceptual studies point to about a factor $\times 2$ increase in resolving power for SCXM compared to CDI, in similar conditions to ours. The present simulation work just comes to support the fact that a complete 3d scan would be realizable in about 24~h time, under realistic assumptions on the experimental setup, detector response and counting algorithms.

\section*{Funding Information}
A. Sa\'a Hern\'andez is funded through the project ED431F 2017/10 (Xunta de Galicia) and D. Gonz\'alez-D\'iaz through the Ramon y Cajal program, contract RYC-2015-18820. C.D.R. Azevedo is supported by Portuguese national funds (OE), through FCT - Funda\c{c}\~{a}o para a Ci\^{e}ncia e a Tecnologia, I.P., in the scope of the Law 57/2017, of July 19.

\section*{Acknowledgments}
We thank Ben Jones and David Nygren (University of Texas at Arlington), as well as our RD51 colleagues for stimulating discussions and encouragement, and specially to David Jos\'e Fern\'andez, Pablo Amedo, and Pablo Ameijeiras for discussions on the K-means method, and Dami\'an Garc\'ia Castro for performing the Magboltz simulations.

\begin{appendices}

\section{Relation between resolution and scan time}~\label{appendixA}
\subsection{Proportional (ideal) case}
We start from the imaging criterion, applied to an arbitrary position of the step motor within a cell-scan:
\begin{equation}
\frac{|N_{f} - N_{0}|}{\sqrt{\sigma_{N_{f}}^2 + \sigma_{N_{0}}^2}} = R   \label{Rose_App}
\end{equation}
where $R=5$ corresponds to the Rose condition. N$_f$ is the number of scattered photons from a water medium with a `to-be-resolved' feature inside it, and N$_0$ contains only water, instead (see Fig.~\ref{fig:Dose}-top). This equation can be re-expressed as:
\begin{equation}
\frac{|N_{f} - N_{0}|}{\sqrt{N_f^2\left(\frac{\sigma_{N_{f}}}{N_{f}}\right)^2 + N_0^2\left(\frac{\sigma_{N_{0}}}{N_{0}}\right)^2}} = R   \label{Rose_App2}
\end{equation}
that, under the assumption $N_{f} \gtrsim N_{0}$, and defining the signal to noise ratio as $S/N \equiv N_f/\sigma_{N_f} \simeq N_0/\sigma_{N_0}$ can be rewritten, in general, as:
\begin{equation}
\frac{1}{\sqrt{2}} \frac{N_{f} - N_{0}}{N_{0}} \times S/N = R  \label{Rose_App3}
\end{equation}

When considering photon counting, it is understood that a relation can be established between the distribution of ionization clusters that are counted in the detector (mean $n$, standard deviation $\sigma_n$) and the distribution of scattered photons (mean $N_f\simeq N_0$, standard deviation $\sigma_{N_f} \simeq \sigma_{N_0}$). If resorting to an unbiased counting algorithm, this relation will be proportional. In that case, the pre-factors on the left-hand-side of eq. \ref{Rose_App3} remain, and any detector-related effect is contained in the quantity:
\begin{equation}
S/N = \frac{N_f}{\sigma_{N_f}} \simeq \frac{N_0}{\sigma_{N_0}} \rightarrow \frac{n}{\sigma_n}
\end{equation}

At fixed number of scattered photons ($\simeq N_0$) the relative fluctuations in the number of counted clusters will increase due to efficiency losses, characteristic emission, and re-scatters on the cell itself, air or structural materials, thereby resulting in a loss of signal to noise. It is convenient to normalize this definition to the Poisson limit for a perfect detector:
\begin{equation}
S/N^* = \frac{1}{\sqrt{N_0}}\cdot S/N
\end{equation}
and so the new quantity $S/N^*$ is now defined between $0$ and $1$, with $S/N = n/\sigma_n$ obtained, in the main document, from detailed simulations of the photon propagation through the experimental setup. Substitution of $N_f$ and $N_0$ by physical quantities in eq. \ref{Rose_App3} yields:
\begin{equation}
\frac{1}{\sqrt{2}} \frac{d(\lambda_f^{-1} - \lambda_w^{-1})}{l\lambda_w^{-1} + 2a\lambda_a^{-1}} \times S/N^* \times \sqrt{N_0} = R  \label{Rose_App4}
\end{equation}
with $d$ being the feature size, $l$ the cell dimension, and $\lambda_{f,w,a}$ the mean free paths in the feature, water and air, respectively, as defined in text. 

Now, we make use of the fact that $N_0=r \cdot \Delta{T_{step}} \cdot (l\lambda_w^{-1} + 2a\lambda_a^{-1})$, with $r$ being the beam rate, $\Delta{T_{step}}$ a time step within the scan, and $\Delta{T_{scan}}$ the total time for a 2d scan: $\Delta{T_{scan}} = \left(\frac{l}{d}\right)^2 \cdot \Delta{T_{step}}$. By replacing $N_0$ in the previous equation we obtain:
\begin{equation}
\frac{1}{\sqrt{2}} \frac{d^2(\lambda_f^{-1} - \lambda_w^{-1})}{l(l\lambda_w^{-1} + 2a\lambda_a^{-1})^{1/2}} \times S/N^* \times \sqrt{r \cdot \Delta{T_{scan}}} = R  \label{Rose_App5}
\end{equation}
from which the time needed for a complete 2d scan can be expressed as:
\begin{equation}
\Delta T_{scan} = R^2 \frac{2 l^2}{d^4}\frac{(l\lambda_w^{-1} + 2a\lambda_a^{-1})}{(\lambda_f^{-1} - \lambda_w^{-1})^2} \frac{1}{S/N^{*,2} \cdot r} \label{T_App}
\end{equation}
and, solving for $d$:
\begin{equation}
d = \left(R^2 2 l^2\frac{(l\lambda_w^{-1} + 2a\lambda_a^{-1})}{(\lambda_f^{-1} - \lambda_w^{-1})^2} \frac{1}{S/N^{*,2} \cdot r \cdot \Delta{T}_{scan}}\right)^{1/4} \label{d_App}
\end{equation}
Expression \ref{d_App} can be approximated under the simplifying assumption that $S/N^*$ is mainly limited by Poisson statistics and by the efficiency of the detector (modelled through a simple binomial distribution), disregarding production of secondary particles or re-scatters across structural materials, hence:
\begin{equation}
S/N^* \!=\! \frac{1}{\sqrt{N_0}} \frac{n}{\sigma_{n}} \! \simeq \! \frac{1}{\sqrt{N_0}}  \frac{N_0 \epsilon}{\sqrt{\epsilon^2 N_0 + \epsilon\cdot(1-\epsilon)\cdot N_0}} \! =\! \sqrt{\epsilon} \label{StoNapp}       
\end{equation}
From which it can be seen that detector efficiency and beam rate enter as a product in the denominator in formulas \ref{T_App} and \ref{d_App}. Consequently, detector inefficiency increases the scan time linearly, as intuitively expected.

\subsection{Non proportional case}

We consider now the more realistic case where there is a non-proportional response of the counting algorithm. This is characterized, for the K-means algorithm implemented in text, as a second order polynomial (Fig. 11):
\begin{equation}
n = a + b r + c r^2
\end{equation}
By analogy, if the K-means parameters are optimized for a certain beam rate, r, the response to cell regions causing a different number of scattered photons $N$, relative to the water-only case, will be:
\begin{equation}
n = a + b \frac{N}{N_0} + c \left(\frac{N}{N_0}\right)^2
\end{equation}
and $a(r)$, $b(r)$, $c(r)$ are now rate-dependent.
Eq. \ref{Rose_App3} should be rewritten, accordingly, as:
\begin{equation}
\frac{1}{\sqrt{2}} \frac{n_{f} - n_{0}}{n_{0}} \times S/N = R
\end{equation}
and the relative variation in $n$ becomes:
\begin{equation}
\frac{n_{f} - n_{0}}{n_{0}} = \frac{1}{a+b+c}\left( b\frac{N_f - N_0}{N_0} + c\frac{N_f^2 - N_0^2}{N_0^2} \right)
\end{equation}
that, for $N_f \simeq N_0$, can be re-expressed as:
\begin{equation}
\frac{n_{f} - n_{0}}{n_{0}} = C_l(r) \frac{N_{f} - N_{0}}{N_{0}}
\end{equation}
with $C_l(r) = \frac{b + 2c}{a + b + c}$. Hence, a loss of linearity during the counting process enters linearly in eq. \ref{Rose_App3}. The general expression for the resolvable feature size as a function of the beam rate is, finally, by analogy with eq. \ref{d_App}:
\begin{equation}
d = \left(R^2 2 l^2\frac{(l\lambda_w^{-1} + 2a\lambda_a^{-1})}{(\lambda_f^{-1} - \lambda_w^{-1})^2} \frac{1}{C_l(r)^2 \cdot S/N^{*,2}\cdot r \cdot\Delta{T}_{scan}}\right)^{1/4} \label{d_App_Clin}
\end{equation}
that is the expression used in the main document, for the achievable resolution as a function of the scan time, under a given imaging criterion $R$. 
The detector response enters this final expression in three ways:
\begin{enumerate}
    \item Through the increased fluctuation in the number of detected clusters, relative to the ideal (Poisson) counting limit, characterized through the signal to noise ratio, $S/N^*$.
    \item The non-linearity of the counting algorithm, $C_l$.
    \item The assumed maximum operating rate, $r$, for which the product $C_l^2\cdot r$ reaches a maximum, as for larger rates stops improving the ability to resolve an image.
\end{enumerate}

\section{EL-TPC parameters}~\label{appendixB}
Here we compile the main parameters used for the simulation of the TPC response, together with additional references when needed.

\begin{table}[h]
\centering
\caption{Parameters of the TPC vessel}
\begin{tabular}{p{0.06\textwidth}p{0.04\textwidth}p{0.09\textwidth}p{0.21\textwidth}}
\hline
$R_{i}$ & 5   & cm       & inner radius \\
$R_{o}$ & 25  & cm       & outer radius \\
$L$     & 50  & cm       & length \\
\hline
\end{tabular}
  \label{tab:Gas_parameters}
\end{table}

\begin{table}[h]
\centering
\caption{Main gas parameters (xenon + 0.4\% CH$_4$)}
\begin{tabular}{p{0.06\textwidth}p{0.04\textwidth}p{0.09\textwidth}p{0.21\textwidth}}
\hline
\multicolumn{4}{l} {in the drift/collection region} \\
\hline
$E_c$       & 110   & V/cm        & collection field \\
$V_{cat}$   & -8.5  & kV          & cathode voltage \\
$F$         & 0.15  &             & Fano factor~\cite{Dave_prop} \\
$W_I$       & 22    & eV          & energy to create an e$^-$-ion pair \cite{Dave_prop}\\ 
$D_T^*$     & 0.548 & mm/$\sqrt{\textnormal{cm}}$ & transverse diffusion coefficient \cite{Pyboltz}\\
$D_L^*$     & 1.52  & mm/$\sqrt{\textnormal{cm}}$ & longitudinal diffusion coefficient \cite{Pyboltz}\\
$v_d$       & 5.12  & mm/$\mu{s}$       & drift velocity \cite{Pyboltz} \\
\hline
\multicolumn{4}{l} {in the electroluminescence (EL) region} \\
\hline
$E_{EL}$    & 6    & kV/cm       & EL field \\
$V_{gate}$  & -3   & kV          & voltage at FAT-GEM entrance (`gate') \\
$v_{d,EL}$  & 13.7 & mm/$\mu$s   & drift velocity~\cite{Pyboltz} \\
\hline
\end{tabular}
  \label{tab:TPC_parameters}
\end{table}

\begin{table}[h!]
\centering
\caption{Parameters of the electroluminescent structure}
\begin{tabular}{p{0.06\textwidth}p{0.04\textwidth}p{0.09\textwidth}p{0.21\textwidth}}
\hline
$r_h$       &  3 & mm & hole radius \\
$t$         &  5 & mm & thickness \\
$p_h$         & 10 & mm & hole-to-hole pitch \\ 
$m_{opt}$   & 250  & ph/e/cm & optical gain~\cite{FATGEM} \\
$P_{scin}$  & 0.5  &    & scincillation probability~\cite{Henriques} \\
\hline
\end{tabular}
  \label{tab:FATGEM_parameters}
\end{table}

\begin{table}[h!]
\centering
\caption{Parameters of the readout}
\begin{tabular}{p{0.06\textwidth}p{0.04\textwidth}p{0.09\textwidth}p{0.21\textwidth}}
\hline
p$_{si}$                   & 10   &    & pitch of SiPM matrix \\
$\Delta{T}_{\textnormal{s}}$    & 0.5   & ${\mu}s$    & time sampling / time per slice \\
$\sigma_{t}$         & 7    & ns & temporal width of SiPM signal \cite{Hamamatsu}    \\
$\sigma_G/G$           & 0.1  & & relative spread of single phe charge in SiPM \cite{Hamamatsu} \\
$\Omega_{TPB}$          & 0.3  &    & geometrical acceptance of SiPM after wavelength shifter \\
$QE_{wls}$              & 0.4  &    & quantum efficiency of wavelength shifter~\cite{Gehman}\\ 
$QE_{si}$        & 0.4  &    & quantum efficiency of SiPM~\cite{Hamamatsu} \\
\hline
\end{tabular}
  \label{tab:TPC_parameters}
\end{table}

\end{appendices}

\end{document}